\documentclass[a4paper,11pt]{article}

\pdfoutput=1 

\usepackage{jcappub} 

\usepackage{amsmath}
\usepackage{amssymb}
\usepackage{graphicx,caption,subcaption}

\makeatletter
\def\@hangfrom#1{\setbox\@tempboxa\hbox{{#1}}%
      \hangindent 0pt
      \noindent\box\@tempboxa}
\makeatother
\def\un#1{\relax\ifmmode\@@underline#1\else
        $\@@underline{\hbox{#1}}$\relax\fi}
\let\du=\du                     
\def\a{\alpha}

\def\d{\delta}

\def\f{\phi}

\def\m{\mu}
\def\n{\nu}
\def\o{\omega}
\def\p{\pi}

\def\s{\sigma}

\def\z{\zeta}
\def\D{\Delta}

\def\L{\Lambda}

\def\bo{{\raise-.3ex\hbox{\large$\Box$}}}               
\def\pa{\partial}                                       
\def\TH{{\raise.2ex\hbox{$\displaystyle \bigodot$}\mskip-4.7mu \llap H \;}}
\def\face{{\raise.2ex\hbox{$\displaystyle \bigodot$}\mskip-2.2mu \llap {$\ddot
        \smile$}}}                                      

\def\VEV#1{\left\langle #1\right\rangle}        
\def\abs#1{\left| #1\right|}                    
\def\leftrightarrowfill{$\mathsurround=0pt \mathord\leftarrow \mkern-6mu
        \cleaders\hbox{$\mkern-2mu \mathord- \mkern-2mu$}\hfill
        \mkern-6mu \mathord\rightarrow$}
\def\dvec#1{\vbox{\ialign{##\crcr
        \leftrightarrowfill\crcr\noalign{\kern-1pt\nointerlineskip}
        $\hfil\displaystyle{#1}\hfil$\crcr}}}           
\def\frac#1#2{{\textstyle{#1\over\vphantom2\smash{\raise.20ex
        \hbox{$\scriptstyle{#2}$}}}}}                   
\def\sfrac#1#2{{\vphantom1\smash{\lower.5ex\hbox{\small$#1$}}\over
        \vphantom1\smash{\raise.4ex\hbox{\small$#2$}}}} 
\def\bfrac#1#2{{\vphantom1\smash{\lower.5ex\hbox{$#1$}}\over
        \vphantom1\smash{\raise.3ex\hbox{$#2$}}}}       
\def\afrac#1#2{{\vphantom1\smash{\lower.5ex\hbox{$#1$}}\over#2}}    
\def\[{\lfloor{\hskip 0.35pt}\!\!\!\lceil}
\def\]{\rfloor{\hskip 0.35pt}\!\!\!\rceil}

\def\du#1#2{_{#1}{}^{#2}}

\def\fracm#1#2{\hbox{\large{${\frac{{#1}}{{#2}}}$}}}

\def\un{\underline}
\def\fracmm#1#2{{{#1}\over{#2}}}

\def\low#1{{\raise -3pt\hbox{${\hskip 0.75pt}\!_{#1}$}}}

\newskip\humongous \humongous=0pt plus 1000pt minus 1000pt
\def\caja{\mathsurround=0pt}
\def\eqalign#1{\,\vcenter{\openup2\jot \caja
        \ialign{\strut \hfil$\displaystyle{##}$&$
        \displaystyle{{}##}$\hfil\crcr#1\crcr}}\,}
\newif\ifdtup

\def\({\left(}
\def\){\right)}

\def\beq{\begin{equation}}
\def\eeq{\end{equation}}
\def\bea{\begin{eqnarray}}
\def\eea{\end{eqnarray}}
\newcommand{\be}{\begin{equation}}
\newcommand{\ee}{\end{equation}}
\newcommand{\nbe}{\begin{equation*}}
\newcommand{\nee}{\end{equation*}}

\newcommand{\lb}{\label}

\title{\boldmath Analytic extensions of Starobinsky model of inflation}

\author[a]{Vsevolod R.~Ivanov,}
\author[b,c,d]{Sergei V.~Ketov,}
\author[e]{Ekaterina O.~Pozdeeva,}
\author[e,1]{Sergey Yu.~Vernov\note{The corresponding author.}}

\affiliation[a]{Physics Department, Lomonosov Moscow State University\\
Leninskie Gory~1, 119991 Moscow, Russia}
\affiliation[b]{Institut f\"ur Theoretische Physik and Riemann Center for Geometry and Physics\\
Leibniz Universit\"at Hannover, Appelstrasse 2, 30167 Hannover, Germany}
\affiliation[c]{Department of Physics, Tokyo Metropolitan University\\
1-1 Minami-ohsawa, Hachioji-shi, Tokyo 192-0397, Japan}
\affiliation[d]{Kavli Institute for the Physics and Mathematics of the Universe (WPI)
\\The University of Tokyo Institutes for Advanced Study, Kashiwa 277-8583, Japan}
\affiliation[e]{Skobeltsyn Institute of Nuclear Physics, Lomonosov Moscow State University,\\
Leninskiye Gory~1, 119991 Moscow, Russia}

\emailAdd{vsvd.ivanov@gmail.com}
\emailAdd{ketov@tmu.ac.jp}
\emailAdd{pozdeeva@www-hep.sinp.msu.ru}
\emailAdd{svernov@theory.sinp.msu.ru}

\keywords{inflation, modified gravity, cosmology}

\arxivnumber{2111.09058}

\abstract{We study several extensions of the Starobinsky model of inflation, which obey all observational constraints on the inflationary parameters, by demanding that both the inflaton scalar potential in the Einstein frame and the $F(R)$ gravity function in the Jordan frame have the explicit dependence upon fields {\it and} parameters in terms of elementary functions. Our models are continuously connected to the original Starobinsky model via changing the parameters. We modify the Starobinsky  $(R+R^2)$ model by adding an $R^3$-term, an $R^4$-term, and an $R^{3/2}$-term, respectively, and calculate the scalar potentials, the inflationary observables and the allowed limits on the deformation parameters by using the latest observational bounds. We find that the tensor-to-scalar ratio in the Starobinsky model modified by the $R^{3/2}$-term significantly increases with raising the parameter in front of that term. On the other side, we deform the scalar potential of the Starobinsky model in the Einstein frame in powers of $y=\exp\left(-\sqrt{\fracmm{2}{3}}\phi/M_{Pl}\right)$, where $\phi$ is the canonical inflaton (scalaron) field, calculate the corresponding $F(R)$ gravity functions in the two new cases, and find the restrictions on the deformation parameters in the lowest orders with respect to the variable $y$ that is physically small during slow-roll inflation.
}

\begin{document}
\maketitle
\flushbottom

\section{Introduction}

The {\it duality} relation between modified $F(R)$ gravity theories and scalar-tensor gravity theories is the standard tool in modern cosmology, see Refs.~\cite{Whitt:1984pd,Maeda:1987xf,Maeda:1988ab,Barrow:1988xh,Rodrigues:2011zi} for the original papers about the duality transformation  and
Refs.~\cite{Sotiriou:2008rp,DeFelice:2010aj,Capozziello:2011et,Ketov:2012yz} for a review of $F(R)$ gravity theories
and their physical applications. In the literature, the duality relation is usually used only in one direction, from an  $F(R)$ gravity model to the equivalent scalar-tensor (or quintessence) model in the Einstein frame with a propagating scalar field. In the context of inflationary models~\cite{Starobinsky:1980te,Saidov:2010wx,Ketov:2012se,Sebastiani:2013eqa,Motohashi:2014tra,Bamba:2015uma,Miranda:2017juz,Motohashi:2017vdc,Cheong:2020rao,Rodrigues-da-Silva:2021jab}, the scalar field is identified with the inflaton having the clear gravitational origin as a physical excitation of the higher-derivative gravity (called scalaron).

A well-known example of the correspondence is given by the celebrated Starobinsky model of inflation
\cite{Starobinsky:1980te}, whose action is given by
\be \lb{starm}
S_{\rm Star.}[g^J_{\m\n}] = \fracmm{M^2_{Pl}}{2}  \int d^4x\sqrt{-g_J} \left( R_J
+ \fracmm{1}{6m^2} R_J^2\right)~,
\ee
where we have introduced the reduced Planck mass $M_{Pl}$ and the inflaton mass $m$, in terms of metric
$g^J_{\m\n}$ having the Ricci scalar curvature $R_J$, with the spacetime signature $(-,+,+,+)$.
The action (\ref{starm}) is dual to the quintessence (or scalar-tensor gravity) action
\be \lb{quint}
S_{\rm quint.} [g_{\m\n},\f] = \int d^4x\sqrt{-g} \left[ \fracmm{M^2_{Pl}}{2}R -\fracmm{1}{2} g^{\m\n}\pa_{\m}\f
\pa_{\nu}\f - V_{\rm Star.}(\f)\right]
\ee
in terms of the canonical scalar $\f$  and another metric $g_{\m\n}$ in the Einstein frame, related to $g^J_{\m\n}$
(in the Jordan frame) by a Weyl transform (see Sec.~2 for details), and having the Ricci scalar $R$. The induced
scalar potential is given by
\be \lb{starp}
V_{\rm Star.}(\f) = \fracmm{3}{4}M^2_{Pl}m^2 \left[ 1- \exp\left(-\sqrt{\fracmm{2}{3}}\f/M_{Pl}\right)\right]^2~.
\ee
The Starobinsky model is known as the excellent model of large-field slow-roll cosmological inflation with very good agreement to the Planck measurements of the Cosmic Microwave Background (CMB) radiation \cite{Planck:2018jri}. In particular, the observable CMB amplitude fixes the only parameter of the Starobinsky model as $m\sim 10^{-5}M_{Pl}$. It is also remarkable that the Lagrangian (\ref{starm}) and the corresponding scalar potential (\ref{starp}) are very simple.

The Starobinsky model (\ref{starm}) is the particular case of the modified  $F(R)$ gravity theories having the action
\begin{equation}
\label{actionFR}
    S_F[g^J_{\m\n}]  =\int d^4 x \sqrt{-g^J}F(R_J)
\end{equation}
with a differentiable function $F$.  In contrast to Starobinsky's Lagrangian quadratic in $R_J$, the $F(R_J)$ theories with generic functions $F$ do not lead to a simple scalar potential, and the duality transformation itself is non-trivial. This is the reason why the induced scalar potentials are often obtained and studied in the $F(R)$ gravity literature by numerical methods. Moreover, when starting from a generic $F$-function, one often arrives at a multi-valued scalar potential with a number of branching cuts and points, see e.g., Ref.~\cite{Saidov:2010wx} for some examples.  Therefore, mathematically and physically well-defined $F(R)$ gravity functions have to be carefully chosen.

One also has to avoid graviton as a ghost and scalaron (inflaton) as a tachyon. It leads to further restrictions
\be \lb{restr2}
\fracmm{dF}{dR_J} >0 \quad {\rm and} \quad \fracmm{d^2F}{dR^2_J} >0
\ee
that restrict possible values of parameters and $R_J$ \cite{Starobinsky:2007hu,Appleby:2009uf}. For example, in the Starobinsky model, the first condition in (\ref{restr2})  is violated for the large {\it negative} $R_J < - 3m^2$, where the duality transformation does not exist.\footnote{An absence of the duality transformation does not imply non-existence of smooth solutions on which the function $dF/dR_J$ changes its sign. In the Friedmann-Lemaitre-Robertson-Walker (FLRW) universe such solutions do exist~\cite{Ivanov:2021ily} but in more general spacetimes, e.g., with a Bianchi I metric, anisotropic instabilities arise and isotropic solutions are not stable~\cite{Figueiro:2009mm}.}

When considering an inflationary model in modified gravity as the effective gravitational theory, its Hubble    function $H$ must be negligible against the Planck scale $M_{Pl}$ and the UV-cutoff  $\L_{UV}$.  This guarantees decoupling of heavy modes (like Kaluza-Klein modes and string theory massive modes),  which is required for consistency~\cite{Dvali:2020cgt}.  In the Starobinsky model, both conditions are satisfied because
 $H\sim {\cal O}(m)$ and $\L_{UV} = M_{Pl}$, while the latter easily follows from expanding the scalar potential
 (\ref{starp}) in power series with respect to the inflaton field $\phi$. In this paper, we confine ourselves to non-negative values of the Ricci scalar curvature well below ${\cal O}(M^2_{Pl})$.

When using an $F(R)$ gravity model for describing inflation, an agreement with the CMB observables (the amplitude of fluctuations $A_s$, the scalar power spectrum index $n_s$ and the
tensor-to-scalar ratio $r$) is also required. As regards the scalar potential, it is reasonable to demand
a Minkowski (or de Sitter) vacuum and the boundedness of the potential from below. Taken together, all these
restrictions also significantly restrict possible choices of $F$ function and scalar potential.

In this paper, we consider analytic deformations of the Starobinsky model (\ref{starm}), which obey all the above mentioned  restrictions, at least for small values of the deformation parameters.
The inverse duality transformation relating a generic scalar potential $V(\f)$ in the action (\ref{quint}) to the corresponding $F(R)$ gravity function in Eq.~(\ref{actionFR}) is known in the parametric form \cite{Motohashi:2017vdc,Ketov:2014kya,Vernov:2019ubo}. However, it is often impossible to get an explicit analytic solution to the $F$-function in terms of elementary functions for a given scalar potential. For example, a simple quadratic scalar potential (a mass term) leads to the complicated $F$-function in terms of the special (Lambert) function. It is therefore of interest to find other cases (generalizing the Starobinsky model) where {\it both} functions $V(\f)$ and $F(R)$ exist in terms of elementary functions. It is certainly relevant for studying the parameter
spaces of cosmological models because analytical methods are superior to numerical methods there.

 The Starobinsky model is just the simplest model of inflation with sharp predictions for the inflationary observables and no free parameters, i.e. it has the maximal predictive power. However, its viable extensions may be required by future experiments, should the observed values of the scalar perturbation index $n_s$ and the tensor-to-scalar ratio $r$ deviate from their values in the Starobinsky model of inflation. Having more freedom in the choice of new parameters consistent with inflation is also useful for other purposes, such as studies of reheating and astrophysical constraints~\cite{Sotiriou:2008rp,DeFelice:2010aj,Capozziello:2011et,Ketov:2012yz}, or formation of primordial black holes \cite{Garcia-Bellido:2017mdw}.  In the context of inflationary models, we are only interested in the physically viable theories consistent with CMB measurements. In this paper, we find some continuous deformations of the Starobinsky model under the additional condition that both functions $V(\f)$ and $F(R)$ can be explicitly given in a finite form in terms of elementary functions, and focus on the deformations with  only one free parameter.

Our motivation can thus be summarized as follows. The Starobinsky model gives sharp predictions
for observables, so that any new viable extension of the model is worth investigating
because future measurements may deviate from its predictions. Also, the Starobinsky model
has no free parameters, whereas its extensions have new parameters. A full investigation of the
parameter space requires both analytic and numeric methods because numerical calculations alone
are often possible only for specific values of the parameters. Given extra parameters, it is important
to provide specific constraints on their possible values. On the technical side, it is desirable
to identify a "small field" in the Starobinsky model, which can serve for an expansion with respect to that
field in possible extensions.

Our paper is organized as follows. In Sec.~2 we review the duality transformations between $F(R)$ gravity theories and scalar-tensor theories in {\it both} directions, formulate the equations of motion, and define the slow-roll approximation for inflation. The polynomial deformations of the Starobinsky model in $F(R)$ gravity by adding the $R^3$ term or the $R^4$ term in the context of slow-roll inflation are studied in Sec.~3. These terms may arise as the quantum gravity corrections from Planck scale physics. We estimate their contribution to inflation by finding the upper limits of their coefficients, via demanding consistency of slow-roll inflation with CMB observations~\cite{Planck:2018jri,BICEPKeck:2021gln}. The scalar potentials in these models are drastically different from that in Eq.~(\ref{starp}) at very large values of the scalar curvature, independently upon the smallness of the deformation parameters. In Sec.~4 we study the impact of adding the $R^{3/2}$ term to the Starobinsky model of inflation, where our methods also apply. In Sec.~5 we begin with the scalar potential (\ref{starp}), give the two new examples of its one-parametric deformation, which are also consistent with observations, and find the corresponding $F(R)$ gravity functions in the explicit analytic  form. Section~6 is our conclusion.

\section{Setup}

\subsection{Duality transformations}

The $F(R)$ gravity action (\ref{actionFR}) can be rewritten as
\begin{equation}
    S_{J} [g^J_{\m\n},\s]=\int d^4 x \sqrt{-g^J} \left[ F_{,\s} (R_J-\sigma)+ F\right]~,
\end{equation}
where the new scalar field $\s$ has been introduced, and $F_{,\s}(\s)=\fracmm{dF(\sigma)}{d\sigma}$~. Eliminating $\s$ via its algebraic equation of motion,
 $R_J=\s$, yields back the action (\ref{actionFR}) when assuming that $F_{,\s\s}(\s)=d^2 F/d\s^2\neq 0$.
After the Weyl transformation of the metric
\be \lb{wtrans}
g_{\mu\nu}=\fracmm{2F_{,\s}(\sigma)}{M^2_{Pl}}g^J_{\mu\nu}
\ee
one gets the following action in the Einstein frame~\cite{Maeda:1988ab}:
\begin{equation}
\label{SE}
S_{E} [g_{\m\n},\s] =\int d^4x\sqrt{-g}\left[\fracmm{M^2_{Pl}}{2}R-\fracmm{h(\sigma)}{2}{g^{\mu\nu}}\partial_\mu{\sigma}\partial_\nu{\sigma}-V\right],
\end{equation}
where we have introduced the functions
\begin{equation}
\label{Ve}
h(\sigma)=\fracmm{3M^2_{Pl}}{2F_{,\s}^2}F_{,\s\s}^2 \quad {\rm and}\quad
V(\s)= M^4_{Pl}\fracmm{F_{,\s} \sigma-F}{4F_{,\s}^2}~.
\end{equation}
Introducing the canonical scalar field $\f$ instead of $\s$ as
\begin{equation}
\label{phidF}
\phi=\sqrt{\fracmm{3}{2}}M_{Pl}\ln\left[\fracmm{2}{M^2_{Pl}}F_{,\s}\right]
\end{equation}
allows one to rewrite the action $S_{E}$  to the standard (quintessence or scalar-tensor) form:
\begin{equation}\label{ActionSe}
    S_E [g_{\m\n},\phi]=\int  d^4x \sqrt{-g}\left[\fracmm{ M^2_{Pl}}{2}R-\fracmm{1}{2}g^{\m\n}\partial_\mu\phi\partial_\nu\phi-V(\phi)\right].
\end{equation}

The inverse transformation reads as follows~\cite{Motohashi:2017vdc,Ketov:2014kya,Vernov:2019ubo}:
\begin{eqnarray}
R_J &=& \left[\fracmm{\sqrt{6}}{M_{Pl}}V_{,\phi}+\fracmm{4V}{M^2_{Pl}}\right]
\exp\left( \sqrt{\fracmm{2}{3}}\fracmm{\phi}{M_{Pl}}\right)~~,
\label{RV}\\
F &=& \fracmm{M^2_{Pl}}{2}\left[\fracmm{\sqrt{6}}{M_{Pl}}V_{,\phi}+\fracmm{2V}{M^2_{Pl}}
\right] \exp\left( 2\sqrt{\fracmm{2}{3}}\fracmm{\phi}{M_{Pl}}\right)~,
\label{FV}
\end{eqnarray}
where $V_{,\phi}=\fracmm{dV}{d\phi}$, defining the function $F(R_J)$ in the parametric form with the parameter $\phi$.

Being motivated by the potential (\ref{starp}),  we find useful to introduce the non-canonical dimensionless field
\be \lb{ydef}
y \equiv \exp\left( -\sqrt{\fracmm{2}{3}}\fracmm{\phi}{M_{Pl}}\right) =\fracmm{M^2_{Pl}}{2F_{,\sigma}} > 0
\ee
because it is (physically) {\it small\/} during slow-roll inflation. Defining $\tilde{V}(y)=V(\phi)$ and using
\begin{equation*}
\fracmm{dV}{d\phi}={}-\sqrt{\fracmm{2}{3}}\fracmm{y}{M_{Pl}}\fracmm{d\tilde{V}}{dy}~,
\end{equation*}
we simplify Eqs.~(\ref{RV}) and (\ref{FV}) as follows:
\begin{equation}
\label{Ry}
R_J=\fracmm{2}{M_{Pl}^2}\left(2\fracmm{\tilde{V}}{y} -  \tilde{V}_{,y}\right),
\end{equation}
\begin{equation}
\label{Fy}
    F=\fracmm{\tilde{V}}{y^2}- \fracmm{\tilde{V}_{,y}}{y} \,,
\end{equation}
respectively.

Equation (\ref{ydef}) can be obtained as a consequence of Eqs.~(\ref{Ry}) and (\ref{Fy}). Using Eqs.~(\ref{ydef}) and (\ref{Ry}), we get
\begin{equation}\label{Fss}
    F_{,\s\s}=\fracmm{M_{Pl}^4}{4\left(2y\tilde{V}_{,y}-2\tilde{V}-y^2\tilde{V}_{,yy}\right)}~~.
\end{equation}

In the Starobinsky model we have
\be \lb{starpd}
\tilde{V}_{\rm Star.}(y)= V_0(1-y)^2~,\quad {\rm where} \quad V_0=\fracm{3}{4}m^2M_{Pl}^2~~.
\ee

The general equations in this Section should be supplemented by demanding the existence of real solutions, choosing appropriate branches and imposing the physical no-ghost and no-tachyon conditions, which restrict choices of the allowed functions $F(R_J)$ and $V(\phi)$.

The simple extensions of the Starobinsky model in the form (\ref{starp}) are given by the so-called {\it T-models} \cite{Lyth:1998xn} or the $\a=1$ {\it attractors} \cite{Kallosh:2013yoa} with the canonical potential
\be \lb{tmod}
V(\f) = f^2\left(\tanh \fracmm{\f}{\sqrt{6}M_{Pl}}\right)
\ee
in terms  of a regular (monotonic) function $f(z)$, where we have introduced the new dimensionless variable
\be \lb{2var}
 z = \tanh \fracmm{\f}{\sqrt{6}M_{Pl}} \geqslant 0~~.
\ee
In terms of this variable, the Starobinsky potential (\ref{starp}) takes the simple form
\be \lb{starp2}
 V_{\rm Star.}(z)= 4V_0\left(\fracmm{z}{z+1}\right)^2~~.
\ee
The simplest T-model of inflation is defined by the even simpler function
\be \lb{stmod}
V(z) = 4V_0 z^2~.
\ee
All these models have the same values of the inflationary observables $n_s$ and $r$ because for large values of
the inflaton field $\f$ we have $z\approx 1-2\exp\left(-\sqrt{2/3}\f/M_{Pl}\right)$.
The inverse transformation (\ref{RV}) and (\ref{FV}) in terms of the new variable (\ref{2var}) takes the form
 \be \lb{RVz}
 R_J =\fracmm{1}{M_{Pl}^2} (1+z) \left[ (1+z) \fracmm{dV}{dz} + \fracmm{4V}{1-z} \right]
 \ee
and
\be \lb{FVz}
F = \fracmm{(1+z)^2}{2}  \left[ \left( \fracmm{1+z}{1-z} \right)  \fracmm{dV}{dz} + \fracmm{2V}{(1-z)^2} \right]~.
\ee
These equations are suitable in the framework of the {\it pole} inflation near $z\approx 1_-$
\cite{Broy:2015qna,Terada:2016nqg,Pallis:2021lwk}. For example, in the Starobinsky model we find
\begin{equation}
\label{starz}
    R_J=6m^2 \fracmm{z}{1-z} \quad {\rm and} \quad
    F_{\rm Star.}=4V_0\fracmm{z}{(z-1)^2}~.
\end{equation}

The variable $z$ is simply connected to the variable $y$ of Eq.~(\ref{ydef}) as
\be \lb{tpar}
z =\fracmm{1-y}{1+y} \quad {\rm and} \quad y = \fracmm{1-z}{1+z}~~.
\ee
It is to be compared to the map $z(\tilde{z})$ between a disc and a half-plane in complex analysis,
\be \lb{tparz}
z =\fracmm{\tilde{z}-1}{\tilde{z}+1} \quad {\rm and} \quad \tilde{z} = \fracmm{1+z}{1-z}~~,
\ee
with $0<z<1$ and $1<t<\infty$ in our (real) case. It gives the mathematical origin of the variable $y$
because $y=\tilde{z}^{-1}$, and allows us to rewrite Eqs.~(\ref{RVz}), (\ref{FVz})  and (\ref{starz})
in the regular form (without poles) in terms of the real variable $\tilde{z}$ that is large during slow-roll inflation.

\subsection{Equations of motion}

In the spatially flat FLRW universe with the metric
\begin{equation*}
ds^2={}-dt^2+a^2(t)\left(dx^2+dy^2+dz^2\right)\,,
\end{equation*}
the action (\ref{ActionSe}) leads to the standard system of evolution equations:
\begin{equation}\label{equ00}
    6M_{Pl}^2H^2={\dot{\phi}}^2+2V,
\end{equation}
\begin{equation}\label{equ11}
    2M_{Pl}^2\dot{H}={}-{\dot{\phi}}^2,
\end{equation}
\begin{equation}\label{equphi}
    \ddot{\phi}+3H\dot{\phi}+V_{,\phi}=0,
\end{equation}
where $H=\dot{a}/a$ is the Hubble parameter, $a(t)$ is the scale factor, and the dots denote the derivatives with respect to the cosmic time $t$. Equation (\ref{equ00}) is the Friedmann equation.
In the inflationary model building, the e-foldings number
\begin{equation}
N_e=\ln \left( \fracmm{a_{\rm end}}{a} \right)~,
\end{equation}
where $a_{\rm end}$ is the value of $a$ at the end of inflation, is considered instead of the time variable.
Using the relation ${d}/{dt}={}-H\, {d}/{dN_e}$, one can rewrite Eq.~(\ref{equ00}) as follows:
\begin{equation}\label{equ00N}
    Q=\fracmm{2V}{6M_{Pl}^2-\chi^2}~,
\end{equation}
where $Q\equiv H^2$ and $\chi=\phi'=-\dot{\phi}/H$, and the primes (here and below)
denote the derivatives with respect to $N_e$. Equations~(\ref{equ11}) and (\ref{equphi}) yield the dynamical system of equations:
\begin{equation}
\label{DynSYSN}
Q'=\fracmm{1}{M_{Pl}^2}Q\chi^2~,\quad
\phi'=\chi~,\quad
\chi'=3\chi-\fracmm{1}{2M_{Pl}^2}\chi^3-\fracmm{1}{Q}\fracmm{dV}{d\phi}~.
\end{equation}
Using Eq.~(\ref{equ00N}), we rewrite the last equation as
\begin{equation}
\label{DynSYSphi}
\chi'=3\chi-\fracmm{1}{2M_{Pl}^2}\chi^3-\fracmm{6M_{Pl}^2-\chi^2}{2V}\fracmm{dV}{d\phi}~.
\end{equation}

\subsection{Slow-roll approximation and inflation observables}

We associate the spacetime of our Universe with the Einstein frame. The slow-roll parameters
are defined by \cite{Liddle:1994dx}
\begin{equation}\label{slr1}
\epsilon=\fracmm{M^2_{Pl}}{2}\left(\fracmm{V_{,\phi}}{V}\right)^2=\fracmm{y^2}{3}\left(\fracmm{\tilde{V}_{,y}}{\tilde{V}}\right)^2\,,
\end{equation}
\begin{equation}\label{slr2}
\eta=M^2_{Pl}\left(\fracmm{V_{,\phi\phi}}{V}\right)=\fracmm{2y}{3\tilde{V}}\left(\tilde{V}_{,y}+y\tilde{V}_{,yy}\right)~.
\end{equation}

The scalar spectral index $n_s$ and the tensor-to-scalar ratio $r$ in terms of the slow-roll parameters are given by \cite{Liddle:1994dx}
\begin{equation}\label{nsr}
    n_s=1-6\epsilon+2\eta,\qquad r=16\epsilon~.
\end{equation}

In the slow-roll approximation, we have $\chi'\ll \chi$ and $\chi\ll M_{Pl}$. Hence, the function $\phi(N_e)$ can be found as a solution of the differential equation
\begin{equation}\label{equphislr}
    \chi\equiv\phi'\simeq \fracmm{M_{Pl}^2}{V}  V_{,\phi}
\end{equation}
when demanding that $\epsilon=1$ corresponds to the end of inflation with $a=a_{\mathrm{end}}$.

Equation~(\ref{equphislr}) is equivalent to
\begin{equation}\label{equyslr}
    y'=\fracmm{2y^2\tilde{V}_{,y}}{3\tilde{V}}~.
\end{equation}

Using Eq.~(\ref{equphislr}), we connect $n_s$ and $V'$ as follows:~\footnote{When using the notion of the effective potential~\cite{Pozdeeva:2019agu}, it is possible to get an analogue of Eq.~(\ref{nsdVN}) in more general models with the scalar field non-minimally coupled to the Ricci scalar and the Gauss-Bonnet term~\cite{Pozdeeva:2020apf,Pozdeeva:2021iwc}.}
\begin{equation}
\label{nsdVN}
    n_s=1+\fracmm{d}{dN_e}\left[ \ln\fracmm{d}{dN_e}\left(-\fracmm{1}{V}\right)\right]
\end{equation}
that allows us to reconstruct $V(N_e)$ for a given $n_s(N_e)$.

The main cosmological parameters of inflation are given by the scalar tilt $n_s$ and the tensor-to-scalar ratio $r$,
whose values are constrained by the combined Planck, WMAP and BICEP/Keck observations of CMB as \cite{Planck:2018jri,BICEPKeck:2021gln}
\be \label{PlanckCMB}
n_s=0.9649\pm 0.0042  \quad ({\rm 68\% CL}) \qquad {\rm and} \qquad  r < 0.036 \quad ({\rm 95\% CL})~.
\ee
The theoretical values of these observables are sensitive to the duration of inflation and the initial value of the inflaton field, $\phi_i$. For instance, in the case of the Starobinsky inflation, we find
$$\begin{tabular}{|c|c|c|}
  \hline
  $\phi_i/M_{Pl} $ & $5.2262$ & $5.4971$  \\ \hline
  $n_s$ & $0.961$ & $0.969$ \\ \hline
  $r$ & $0.0043$ &  $0.0027$\\ \hline
 $N_e $ &$49.258$ & $62.335$\\
  \hline
\end{tabular}$$

The values of $n_s$ and $r$ do not depend upon the scalaron mass $m$.  The amplitude of scalar perturbations is given by
\be
\label{As}
A_s=\fracmm{2V}{3\pi^2M_{Pl}^4 r}~~,
\ee
while its observed value (Planck) is $A_s=2.1\times 10^{-9}$. Therefore, Eq.~(\ref{As}) relates the height
of the inflationary potential to the tensor-to-scalar ratio $r$.

In the Starobinsky model, we have
\be \lb{Am}
A_s = \fracmm{N^2_e m^2}{24\p^2 M_{Pl}^2}
\ee
that determines the value of $m/M_{Pl}\sim {\cal O}(10^{-5})$.

The primordial power spectra of scalar and tensor perturbations in the inflationary models based on $f(R)$ gravity
were first quantitatively derived in Refs.~\cite{Starobinsky:1983zz,Hwang:1996xh}.

\section{Polynomial modifications of Starobinsky model}

\subsection{The $\left(R+R^2+R^3\right)$ gravity models of inflation}

To the best of our knowledge, adding the higher-order terms in $R$ was first proposed in Ref.~\cite{Barrow:1988xh}. The slow-roll large-field inflation models, continuously connected to the Starobinsky model, were also studied in Refs.~\cite{Berkin:1990nu,Saidov:2010wx,Huang:2013hsb,Sebastiani:2013eqa,Motohashi:2014tra,Bamba:2015uma,
Odintsov:2016imq,Miranda:2017juz,Cheong:2020rao,Rodrigues-da-Silva:2021jab}. We revisit only those models that allow a fully analytic treatment
 in our approach by using the non-canonical field $y$ defined by Eq.~(\ref{ydef}).
We do not change the coefficient in front of the $R^2$ term but add a single term with a higher power in $R$  and a dimensionless parameter in front of it. The most obvious option is a modification of the Starobinsky model by adding an $R^3$ term, while the corresponding scalar potential can be derived in the analytic form. It is natural to interpret the higher-order terms in $R$ as the quantum gravity corrections to the Starobinsky model. It is our purpose to evaluate the size of those corrections that are consistent with the recent CMB measurements~\cite{Planck:2018jri,BICEPKeck:2021gln} by using our methods.

A generic $(R+R^2+R^3)$ gravity action is given by
\be \lb{starr3}
S_{\rm 3-gen.} = \fracmm{M^2_{Pl}}{2} \int d^4x \sqrt {-g_J} \left[  (1+\delta_1) R_J +
\fracmm{(1+\delta_2)}{6m^2} R^2_J + \fracmm{\delta_3}{36m^4} R^3_J \right]~,
\ee
where we have introduced the three dimensionless parameters $\d_{i}$, $i=1,2,3$. Similarly to the Starobinsky case, inflation is supposed to be mainly driven by the $R^2$ term with the dimensionless coefficient in the action, while the coefficient in front of the additional $R^3$ term has the negative (mass) dimension. The latter is usually exploited in
the inflation literature via the standard argument that the higher-order curvature terms (beyond the quadratic order) are suppressed by powers of the Planck mass and, therefore, are irrelevant. We use the Starobinsky mass in Eq.~(\ref{starr3})  instead, while the variable $R_J/m^2$ is not small during inflation (our parameters $\d_i$ do not have to be small). We also assume that all the coefficients in $S_{\rm 3-gen.}$ are non-negative in order to avoid problems with ghosts and negative values of the scalar potential.

The corresponding inflaton scalar potential (\ref{Ve}) is given by
\be
\lb{3genpot}
V(\s) = \fracmm{M^2_{Pl}\s^2\left(1+\d_2 +\fracmm{\d_3}{3m^2}\s\right)}{12m^2\left[ 1+ \d_1 +\left( \fracmm{1+\d_2}{3}\right) \fracmm{\s}{m^2} + \fracmm{\d_3}{12m^4}\s^2\right]^2}=\fracmm{16V_0\tilde{\s}^2\left[3(1+\d_2)+\d_3\tilde{\s}\right]}{3\left[12(1+\delta_1)+4(1+\delta_2)\tilde{\s}+\d_3\tilde{\s}^2\right]^2}~~,
\ee
where the dimensionless variable $\tilde{\s}=\s/m^2$ has been introduced.

It is easy to see that $V(0)=0$, $V(\tilde{\s})>0$ at $\tilde{\s}>0$, and $V$ tends to zero at $\tilde{\s}\rightarrow+\infty$, while the potential has a maximum at some positive value of $\tilde{\s}$.
The extreme equation $V'=0$ has only one positive root given by
\begin{equation}
\label{DV0}
    \tilde{\s}_{\rm max.}=6\sqrt{ \fracmm{1+\d_1}{\d_3} }~~.
\end{equation}

This case is qualitatively different from the pure $R^2$-gravity inflation, because at $\d_3=0$ the potential $(\ref{3genpot})$
 is a monotonically increasing function of $\tilde{\s}$, approaching a positive constant at $\tilde{\s}\rightarrow+\infty$, whereas
when $\d_3>0$ one must have $\tilde{\s}<\tilde{\s}_{\rm max.}$ during and after (hilltop) inflation.

To study the impact of the $R^3$-term on inflation in more detail, let us consider the simplest non-trivial case with $\d_1=\d_2=0$, in which Eq.~(\ref{ydef}) implies
\be
\label{fprime3}
\fracmm{1}{y}=1 + \fracmm{1}{3}\tilde{\s}+\fracmm{\d_3}{12}\tilde{\s}^2\,.
\ee

Equation~(\ref{fprime3}) is a quadratic equation on $\tilde{\s}$ as a function of $y$. The only positive root of this equation is given by
\be
\label{sroot3}
\tilde{\s} = \fracmm{2}{\d_3}\left[\sqrt{1+3\d_3\left(y^{-1}-1\right)} -1 \right]=\fracmm{2}{\d_3}\left[\sqrt{1+3\d_3\left( \mathrm{e}^{ \sqrt{\fracmm{2}{3}}{\phi/M_{Pl}}}-1\right) } -1 \right]\,.
\ee

Using Eqs.~(\ref{Ve}) and (\ref{ydef}), we find the scalar potential in terms of $y$ or the inflaton field $\phi$ as follows:
\begin{equation}
 \label{pot3}
\begin{split}
\tilde{V}(y)& ~=\fracmm{4V_0}{27\d_3^2y}\left[y+2\sqrt{y(y+3\d_3(1-y))}\right]\left(y-\sqrt{y(y+3\d_3(1-y))}\right)^2,\\
V(\phi)& ~=\fracmm{4V_0}{27\d_3^2}\mathrm{e}^{-2\sqrt{2/3}\,\phi/M_{Pl}}\left(
\sqrt{1+3\d_3\left( \mathrm{e}^{\sqrt{2/3}\,\phi/M_{Pl}}-1\right)}-1\right)^2 \times \\
&  ~\times \left[1+ 2 \sqrt{1+3\d_3\left(\mathrm{e}^{\sqrt{2/3}\,\phi/M_{Pl}}-1\right)} \right]~.
\end{split}
\end{equation}
It is worth noticing that $\tilde{V}_{\rm Star.}(y)$ is reproduced in the limit $\d_3\rightarrow 0$.

When $\d_3=1/3$, the potential (\ref{pot3}) is greatly simplified to
\be
\label{pot3d3}
V(\phi) = \fracmm{4}{3} V_0\left[y^2 -3y + 2\sqrt{y}\right]=\fracmm{4}{3} V_0 \mathrm{e}^{-2\sqrt{2/3}\,\phi/M_{Pl}}
\left( 1 + 2 \mathrm{e}^{ \sqrt{3/2}\,\phi/M_{Pl}} - 3  \mathrm{e}^{\sqrt{2/3}{\,\phi/M_{Pl}}}
\right)~.
\ee

 In terms of the variable $z$ defined by Eq.~(\ref{2var}), the potential reads
\begin{equation}\label{Vz}
    V(z)=\fracmm{2V_0\left[(1+(18\delta_3-1)z)(z-1)^2-\left[(z-1)^2-12\delta_3(z-1)z\right]^{3/2}\right]}{27\delta_3^2(1+z)^2(z-1)}~~.
\end{equation}
We also find in this case
\begin{equation}\label{F3Rz}
    R_J=\fracmm{m^2\left(z-1+\sqrt{(12\delta_3z+1-z)(1-z)}\right)}{\delta_3(1-z)}
\end{equation}
and
\begin{equation}
    z=\fracmm{R_J\left(\delta_3R_J+2m^2\right)}{12m^4+\delta_3R_J^2+2m^2R_J}=\fracmm{\tilde{\s}\left(\delta_3\tilde{\s}+2\right)}{12+\delta_3\tilde{\s}^2+2\tilde{\s}}~~.
\end{equation}

\begin{figure}[htb]
\begin{subfigure}{.4\textwidth}
 \centering
 \includegraphics[width=1\linewidth]{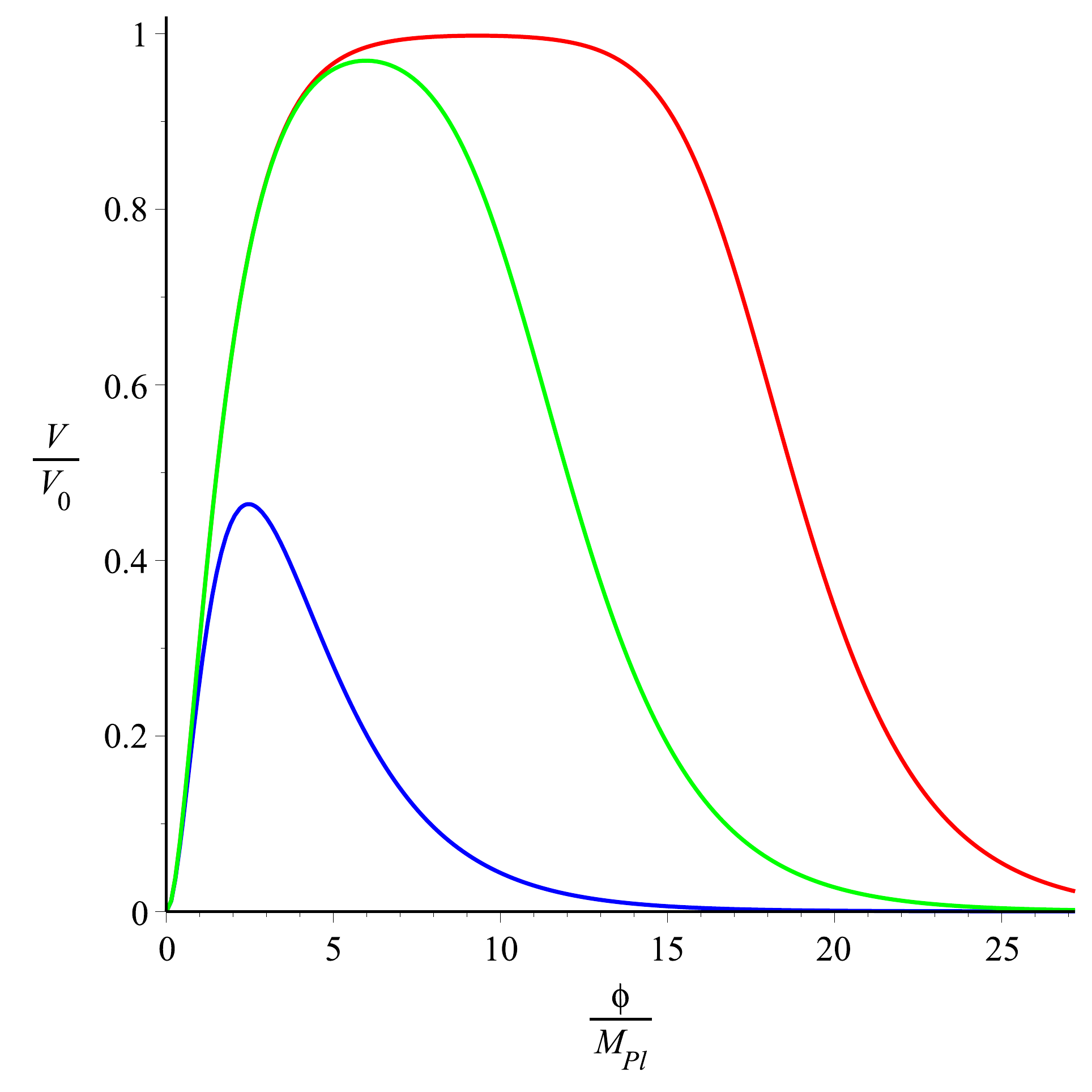}
\end{subfigure}
\captionsetup{width=.9\linewidth}
\caption{The normalized potential $V(\phi)/V_0$ of Eq.~(\ref{pot3}) for $\d_3=0.000001$ (red),  $\d_3=0.000247$ (blue), and  $\d_3=1/3$ (green).
\label{Fig_cubicR}}
\end{figure}

The profile of the potential $V(\phi)$ is given on Fig.~\ref{Fig_cubicR}. The maximum occurs at
\begin{equation}
\label{phimax}
    \phi_{\rm max.}=\sqrt{\fracmm32}\,M_{Pl}\ln\left[2\left(2+\d_3^{-1/2}\right)\right]~,
\end{equation}
while this value does not depend upon $m$.

As is clear from Fig.~\ref{Fig_cubicR}, the plateau of the potential (on the left-hand-side from the maximum) becomes
longer, as well as the duration of slow-roll, with decreasing $\d_3$.

The key discriminator for viable inflation in the given class of models is the value of the scalar perturbation index $n_s$. For example, in the case of the potential (\ref{pot3d3}) with $\d_3=1/3$, the $n_s$ never exceeds $0.9$ for any value of the canonical inflaton field $\f$, so that it is not suitable for inflation.

\begin{figure}[htb]
\begin{subfigure}{.45\textwidth}
 \centering
 \includegraphics[width=1\linewidth]{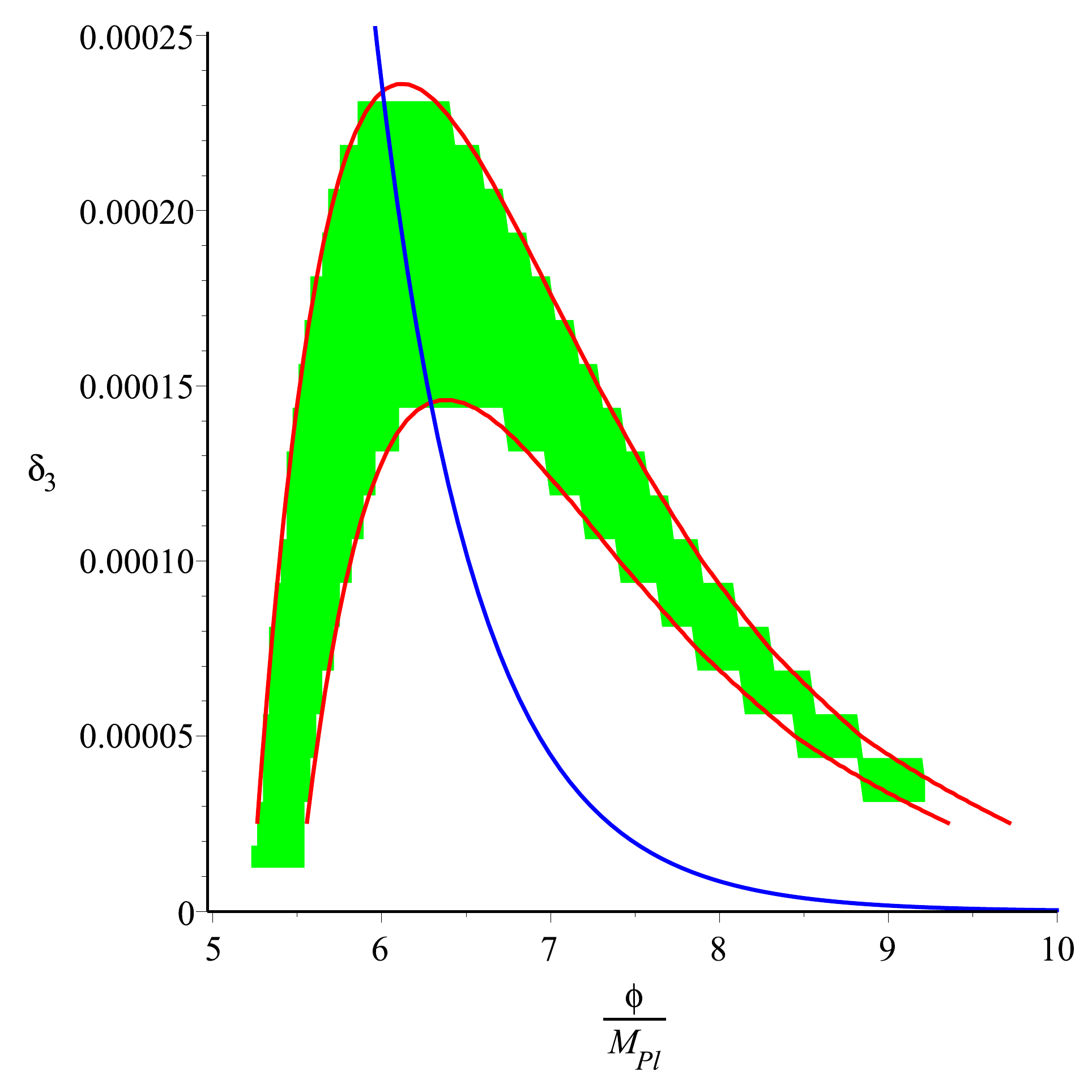}
\end{subfigure}
\begin{subfigure}{.45\textwidth}
  \centering
 \includegraphics[width=1\linewidth]{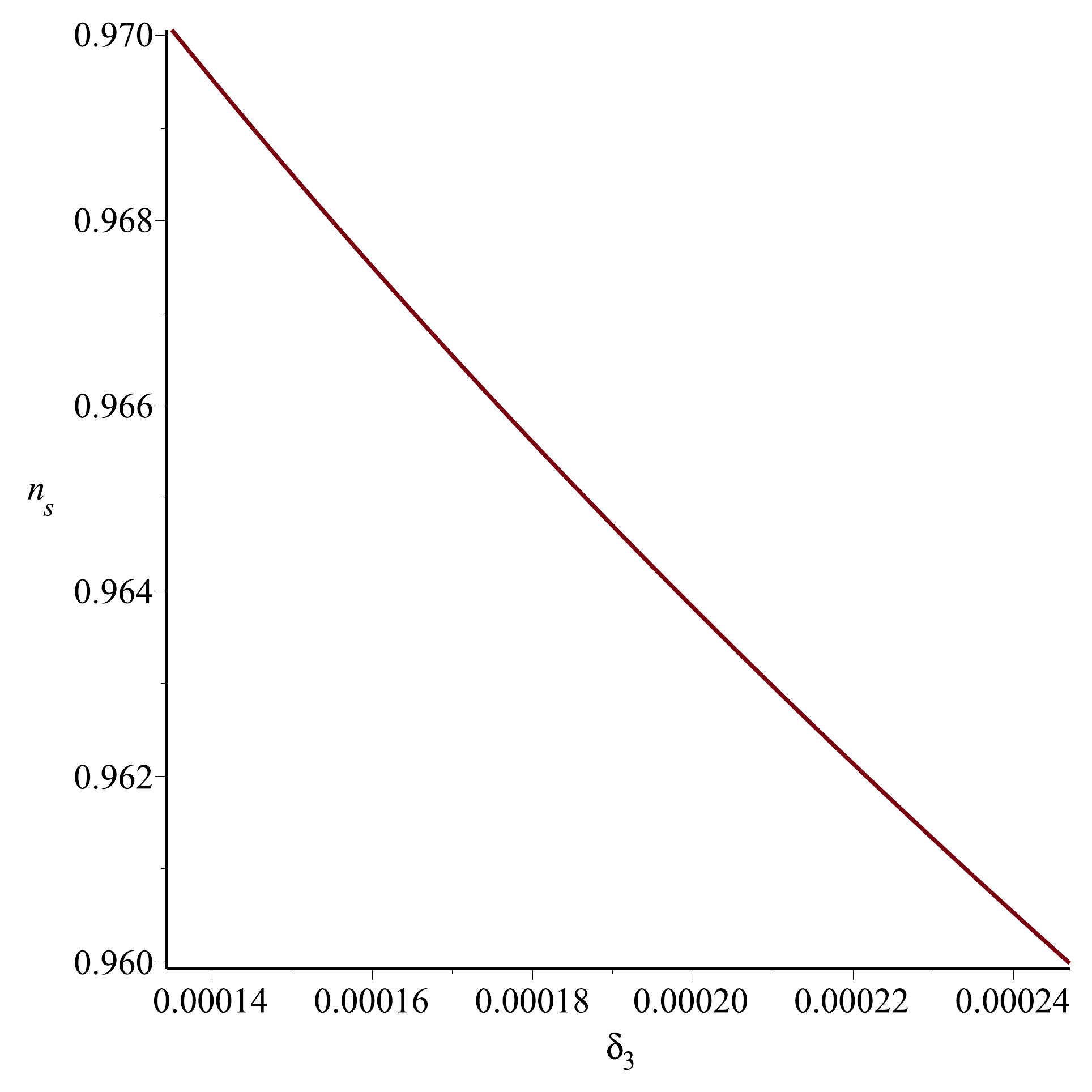}
\end{subfigure}
\captionsetup{width=.9\linewidth}
\caption{The allowed range of $\d_3$ and $\phi$ from the observational constraints (Planck): $0.961<n_s<0.969$ (left), and the dependence of $n_s$ upon $\d_3$ (right), under the assumption that inflation started at the maximum of the potential. }
\label{Fig_cubic_ns}
\end{figure}

The condition $\phi_i<\phi_{\rm max.}$ yields the additional restriction on the possible initial values of $\phi$.  Equation  (\ref{phimax}) can be written to the following form:
\begin{equation}
    \d_3=\fracmm{4}{\left(\mathrm{e}^{\sqrt{2/3}\,\phi_{\rm max.}/M_{Pl}}-4\right)^2}~~,
\end{equation}
being represented by the blue curve on the left-hand-side of Fig.~\ref{Fig_cubic_ns}.

The upper bound on the parameter $\d_3$ can be estimated by assuming the observable value of $n_s$ to be calculated at the maximum of the potential. Then we find
\begin{equation}
n_s(\phi_{\rm max.})=1-\fracmm{8\sqrt{\d_3} \left(1 + 4\sqrt{\d_3} + 4\d_3\right)}{3\left(3\sqrt{\d_3}+1\right)
\left(2\sqrt{\d_3}+1\right)^2}~~.
\end{equation}

Since observations require $n_s>0.960$, we get $\d_3<0.0002467$. The dependence of $n_s$ upon $\d_3$
is given on the right-hand-side of Fig.~\ref{Fig_cubic_ns}. Therefore, the domain of allowed values of $\d_3$ and
$\phi$ is highly restricted.

Since viable inflation requires $\d_3\ll 1$, it allows us to consider $\d_3$ as a truly small parameter and expand  the
potential in power series of $\delta_3$ as follows:
\begin{equation}\label{Vdeltaexpand}
    \tilde{V}(y)=V_0(y-1)^2\left[1+\fracmm{y-1}{2y}\d_3+\fracmm{9(y-1)^2}{16y^2}\d_3^2+{\cal {O}}\left(\d_3^3\right)\right]~.
\end{equation}
In this approximation, using Eqs.~(\ref{slr1}) and (\ref{slr2}), we get
\begin{equation*}
\epsilon\simeq\fracmm{4y^2}{3(y-1)^2}+\fracmm{2}{3(y-1)}\d_3+\fracmm{14y+1}{12y^2}\d_3^2~~,
\end{equation*}
\begin{equation*}
 \eta\simeq\fracmm{4y(2y-1)}{3(y-1)^2}+\fracmm{3y+1}{3y(y-1)}\d_3+\fracmm{21y+16}{12y^2}\d_3^2~~,
\end{equation*}
and, hence,
\begin{equation}\label{ns3}
    n_s\simeq 1-\fracmm{8y(y+1)}{3(y-1)^2}-\fracmm{2(3y-1)}{3y(y-1)}\d_3-\fracmm{21y-13}{6y^2}\d_3^2~~.
\end{equation}

The conditions $\epsilon(y_{end})=1$ and $y_{end}=2\sqrt{3}-3$ at $\d_3=0$ give
\begin{equation}\label{F3end}
    y_{end}=2\sqrt{3}-3+\fracmm{45-26\sqrt{3}}{12\sqrt{3}-21}\d_3+\fracmm{2146\sqrt{3}-3717}{9\left(4\sqrt{3}-7\right)^2}\d_3^2\,,
\end{equation}

Solving Eq.~(\ref{equyslr}), we get for small $\d_3$:
\begin{equation}\label{F3Ney}
    N_e=\fracmm{3(y\ln(y)+1)}{4y}+\fracmm{3y^2-3y+1}{16y^3}\d_3+\fracmm{3(70y^4-110y^3+80y^2-25y+2)}{640y^5}\d_3^2-N_0\,,
\end{equation}
where $N_0$ can be obtained by the condition $N_e(y_{end})=0$:
\begin{equation}\label{F3N0}
\begin{split}
N_0&=\fracmm{(1896102\sqrt{3}-3284145)\ln(2\sqrt{3}-3)-293328\sqrt{3}+508059}{4(2\sqrt{3}-3)^5(4\sqrt{3}-7)^2}+\\
&\fracmm{2578194\sqrt{3}-4465563}{16(2\sqrt{3}-3)^5(4\sqrt{3}-7)^2}\d_3+\fracmm{187776774\sqrt{3}-325238913}{640(2\sqrt{3}-3)^5(4\sqrt{3}-7)^2}\d_3^2.
\end{split}
\end{equation}
With a value of $\d_3$ within $0\leqslant\d_3\leqslant 0.00025$ we obtain $N_0\simeq 1.040$.

In Fig.~\ref{Fig_cubic_ns}, one can see that suitable values of $\phi$ are not less than in the case of the Starobinsky inflation. So, to calculate inflationary parameters we should take $\phi\geqslant 5.226\,M_{Pl}$ that corresponds to $y\leqslant 0.0140$. For these values, $N_e$ is an increasing function of $\d_3$, so the number of e-folding during inflation is always more than $49.26$ that corresponds to the Starobinsky model. The request $N_e<65$ gives the additional restriction of the maximal value of $\d_3$. Namely, for $\d_3=0.00012$, the condition $n_s>0.961$ gives $\phi>5.438 \, M_{Pl}$. The corresponding $N_e>65.1$ that is not suitable for inflationary scenario.

We come to the conclusion that the model under investigation gives the inflationary parameters that do not contradict observations only if $\delta_3<0.00012$ and the inflation started in the narrow domain of the scalar field $\phi$ values (a part of the marked green domain in the left picture of Fig.~\ref{Fig_cubic_ns}), which implies that this inflationary scenario is rather unrealistic. The same model was also studied in detail in Ref.~\cite{Cheong:2020rao,Rodrigues-da-Silva:2021jab}. A similar inflationary model in the framework of the unimodular gravity was proposed in Ref.~\cite{Odintsov:2016imq}.

It is no surprise that the dimensionless parameter $\delta_3$ in front of the $R^3$-term must be very small\footnote{It also applies to the parameter $\delta_4$ in front of the $R^4$-term studied in the next Subsection 3.2.} because slow roll inflation in $f(R)$ gravity only applies for the range of $R$ (in the large curvature regime) where $f(R)/R^2$ is a slowly-changing function of $R$, together with its first and second derivatives with respect to $\ln R$ \cite{Appleby:2009uf}. Our considerations above provide the quantitative estimates for the $\delta_3$ and $n_s$.

\subsection{The $\left(R+R^2+R^4\right)$ gravity model of inflation}

In this Subsection, we consider another model defined by
\begin{equation}
F(R_J)= \fracmm{M^2_{Pl}}{2}\left[ R_J + \fracmm{1}{6m^2} R^2_J +\fracmm{\delta_4 R_J^4}{48 m^6}\right]
\end{equation}
with the dimensionless parameter $\delta_4>0$, as the natural alternative to the previous model.

We compute the inflaton scalar potential from Eq.~(9) as follows:
\begin{equation}\label{2V}
    V=V_0\fracmm{\tilde{\s}^2(8+3\tilde{\s}^2\delta_4)}{72{y}^{-2}}=V_0\fracmm{2\tilde{\sigma}^2\left(8+{3\delta_4 \tilde{\sigma}^2}\right)}{\left(12+4{\tilde{\sigma}}+{\delta_4 {\tilde{\sigma}}^3}\right)^2}~~,
\end{equation}
where $y$ is related to $F'$ by Eq.~(\ref{ydef}). In the case under consideration we have
\begin{equation}\label{ym1}
y^{-1}=1+\fracmm{\tilde{\s}}{3}+\fracmm{\delta_4}{12}\tilde{\s}^3~~.
\end{equation}

Solving the cubic equation (\ref{ym1}) yields $\tilde{\s}$ in terms of $y^{-1}$~,
\be
\lb{cubics}
\tilde{\s}=\fracmm{Z}{3\delta_4}-\fracmm{4}{Z}~,\quad {\rm where} \quad Z=\sqrt[3]{\left(162(y^{-1}-1)+6\sqrt{3}\sqrt{243(y^{-1}-1)^2+16{\delta_4}^{-1}}\right)\delta_4^2}~.
\ee

Assuming $\delta_4\ll 1$ and using Eq.~(\ref{cubics}), we can expand $\tilde{\s}^2$ in power series of $\delta_4$ near zero as follows:
\begin{equation}\tilde{\s}^2 =9(y^{-1}-1)^2-\fracmm{81(y^{-1}-1)^4}{2}\delta_4+\fracmm{5103(y^{-1}-1)^6}{16}\delta_4^2 +{\cal O}\left(\delta_4^3\right)~.
\label{varsigma2}\end{equation}

Substituting Eq.~(\ref{varsigma2}) into Eq.~(\ref{2V}), we get the scalar potential with the same accuracy,
\begin{eqnarray} \lb{approxV}
V(y)&\approx& V_0\left[\fracmm{\left(y^{-1}-1 \right)^{2}}{{y^{-2}}}-\fracmm{9}{8}{\fracmm{\left({y^{-1}}-1\right)^{4}}{y^{-2}}}{\delta_4}+\fracmm{81}{16}\fracmm{\left( {y^{-1}}-1\right)^{6}}{{y^{-2}}}\delta^{2}_4\right]\nonumber\\
&\approx& \tilde{V}_{\rm Star.}(y)\left[1-\fracmm{9}{8}\left(y^{-1}-1\right)^2\delta_4+\fracmm{81}{16}\left(y^{-1}-1\right)^4\delta^2_4\right]~,
\end{eqnarray}
where the potential $\tilde{V}_{\rm Star.}(y)$ has been defined by Eq.~(\ref{starpd}).

\begin{figure}[htb]
 \centering
 \includegraphics[width=0.5\textwidth]{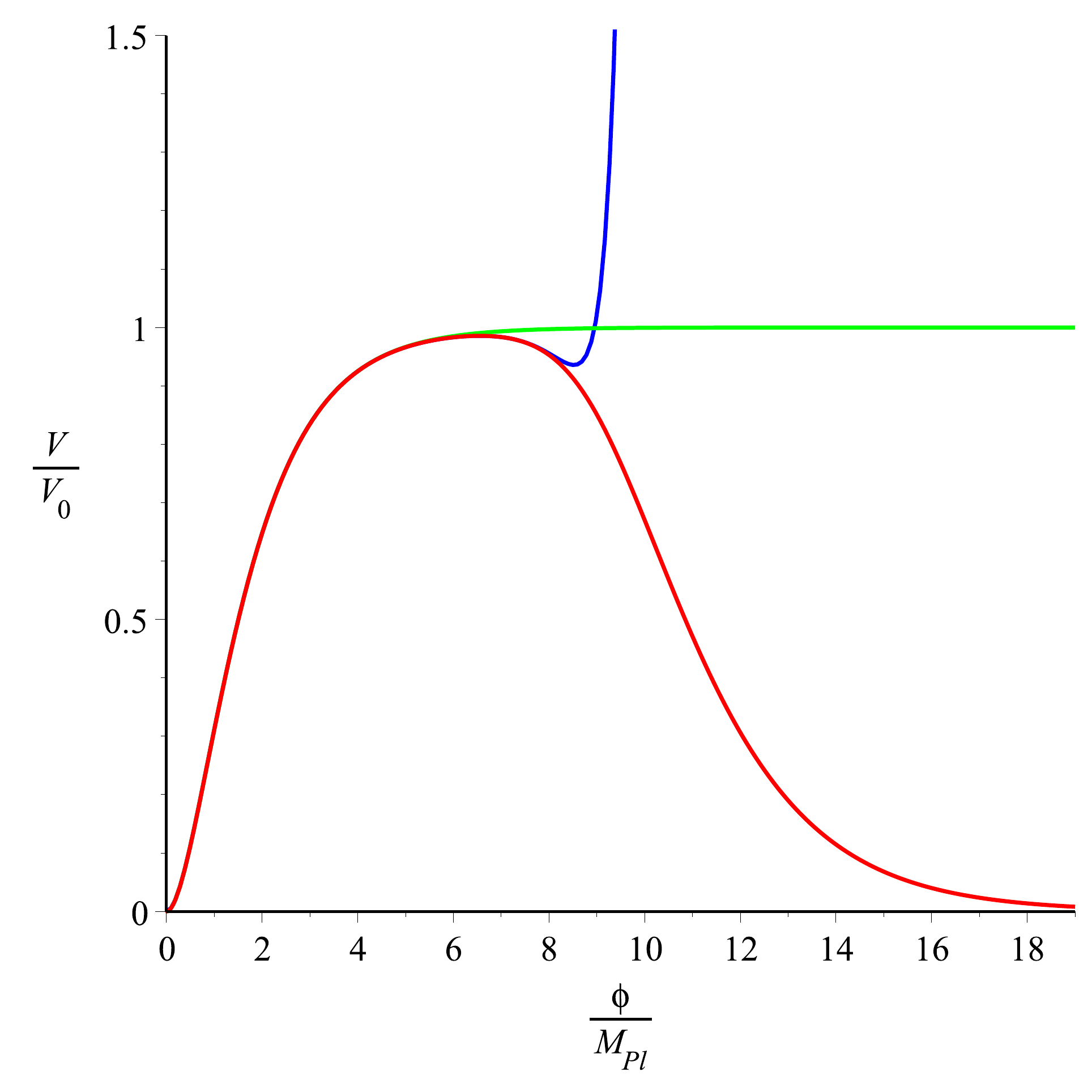}
 \captionsetup{width=.9\linewidth}
\caption{The scalar potential in the $(R+R^2+R^4)$ model (red) versus the scalar potential in the
Starobinsky $(R+R^2)$ model (green). The parameter value is $\delta_4= 10^{-7}$. The approximated
potential (\ref{approxV}) in the $(R+R^2+R^4)$ model is given by the blue line with the same value of
$\delta_4=10^{-7}$.}
\label{Fig_quart}
\end{figure}

The profile of the scalar potential in the $(R+R^2+R^4)$ model versus the scalar potential of the
Starobinsky $(R+R^2)$ model is given in Fig.~\ref{Fig_quart}.

The slow-roll parameters are
\begin{equation}
\epsilon\approx\fracmm{4}{3\left({y}^{-1}-1\right)^2}-3{y}^{-1}\delta_4+ \fracmm{27\left({y}^{-1}+14\right)\left({y}^{-1}-1\right)^2{y}^{-1}}{16}\delta_4^2
\end{equation}
and
\begin{equation}
\eta\approx{}-\fracmm{4\left({y}^{-1}-2\right)}{3\left({y}^{-1}-1\right)^2}-\fracmm{3\left(2{y}^{-1}+3\right){y}^{-1}}{2}\delta_4-\fracmm{81{y}^{-1}
\left(-10{y}^{-3}+13{y}^{-2}+4{y}^{-1}-7\right)}{16}\delta_4^2~.
\end{equation}

Accordingly, the inflationary observables $n_s=1-6\epsilon+2\eta$ and $r=16\epsilon$ are given by
\begin{equation*}
n_s\approx\fracmm{3{y}^{-2}-14{y}^{-1}-5}{3\left({y}^{-1}-1\right)^2}-3{y}^{-1}\left(2{y}^{-1}-3\right)\delta_4-\fracmm{81{y}^{-1}\left(-9{y}^{-3}+25{y}^{-2}-23{y}^{-1}+7\right)\delta_4^2}{8}
\end{equation*}
and
\begin{equation*}
r\approx\fracmm{64}{3\left({y}^{-1}-1\right)^2}-48{y}^{-1}\delta_4+27{y}^{-1}\left({y}^{-1}+14\right)\left({y}^{-1}-1\right)^2\delta_4^2~~.
\end{equation*}

\begin{figure}[htb]
\begin{subfigure}{.49\textwidth}
 \centering
 \includegraphics[width=1\linewidth]{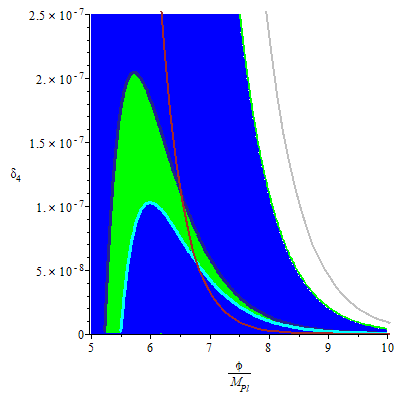}
\end{subfigure}
\begin{subfigure}{.49\textwidth}
\centering
 \includegraphics[width=1\linewidth]{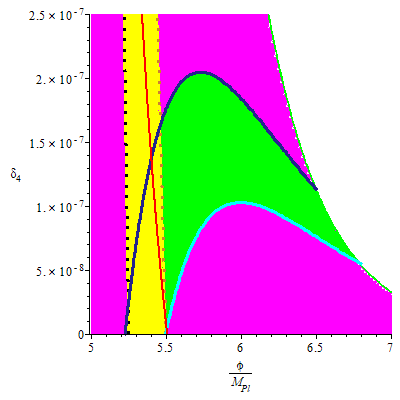}
\end{subfigure}
\caption{The green area on these pictures describes the values of the parameter $\delta_4$ and $\phi=\phi_i$ that are in agreement with the observed values of the spectral index $n_s$ and the tensor-to-scalar ratio $r$.
The upper boundary of the green area (dark blue curve) corresponds to $n_s=0.961$ and the lower boundary (blue curve) corresponds to $n_s=0.969$.
In the left picture, the blue area gives the values of $\delta_4$ and $\phi_i$ leading to the observed values of $r$, the red curve corresponds to the maximum of the potential $V$, and the grey curve shows the boundary of applicability of our approximation. In the right picture, the magenta area shows the values of the parameter $\delta_4$ and $\phi_i$ required for inflation with a decreasing scalar field. The yellow area corresponds to the e-foldings number  $N_e$ of the Starobinsky inflation for the allowed interval $0.961\leqslant n_s\leqslant 0.969$, the left black dotted line corresponds to $N_e\approx 49.258$, the right orange dotted line corresponds to $N_e\approx 62.335$, and the red line corresponds to $r=0.0027$. }
\label{E}
\end{figure}

The e-foldings number $N_e$ as a function of the inflaton field is
\begin{equation}
\begin{split}
N_e &~ \approx \fracmm{3}{4} \left( y^{-1} + \ln y \right)
+\fracmm{27}{128} \left({y}^{-1}-1\right)^4\delta_4 \\
&~ +\fracmm{243}{256} \left({7}^{-1}{{y}^{-7}}-2{y}^{-6}+9{y}^{-5}-20{y}^{-4}+25{y}^{-3}-18{y}^{-2}+7{y}^{-1}\right)\delta_4^2-N_0~~,
\end{split}
\end{equation}
where the integration constant $N_0$ is fixed by the condition $\epsilon(y_{end})=1$. A numerical solution to the condition $\epsilon=1$ with
$\delta_4\leqslant 10^{-6}$ yields the approximate values of the field and the e-foldings number at the end of inflation
as $\phi_{end}\approx0.9402M_{Pl}$ and $N_0\approx1.040$. The relevant values of the canonical inflaton field $\phi_{i}$
at the beginning of inflation versus the values of $\delta_4$ are given in Fig.~\ref{E}.

The red curve in the middle of the left picture in Fig.~\ref{E} gives the function $\delta(\phi_{\max})$, where $\phi_{\max}$ is the point of the maximum potential of $V$. This function can be presented in an analytical form
 \begin{equation*}
 {\delta_{4}}(\phi_{max})=\fracmm{{y}_{max}^{-1}+1-\sqrt{{y}_{max}^{-2}-30{y}_{max}^{-1}-15}}{9\left(2{y}_{max}^{-1}+1\right)\left({y}_{max}^{-1}-1\right)^2}~~,
\end{equation*}
where ${y}_{max}=y(\phi_{max})$.
The scalar field $\phi$ tends to zero during inflation if its initial value is less than $\phi_{\max}$.

As shown in Fig.~\ref{E}, the maximum allowed value of $\delta_4$ is about
$2\times 10^{-7}$. In the case of the Starobinsky inflation, $\delta_4=0$,
limiting the maximum value of the scalar spectral index $n_s$ leads to a minimum value of the ratio of the tensor to the scalar $r=0.0027$. When $\delta_4>0$, the minimum value of $r$ decreases.
 In the right picture of Fig.~\ref{E}, the red line corresponds to the value $r=0.0027$ for $\delta_4\neq0$. To the left of the red line we have $r>0.0027$, and  $r<0.0027$ to the right.

Our consideration provides quantitative estimation for value of $\delta_4$ for which the parameters $n_s$ and $r$ are consistent with current observations.

\section{The $\left(R+R^{3/2}+ R^2\right)$ gravity model of inflation}

\subsection{The inflaton potential}

Our methods apply to yet another model of inflation based on the modified gravity with the $(R+R^{3/2}+ R^2)$ terms,
\begin{equation}
\label{FR32}
    F(R_J)=\fracmm{M^2_{Pl}}{2}\left[ R_J + \fracmm{1}{6m^2} R^2_J + \fracmm{\d}{m} R^{3/2}_J \right]~,
\end{equation}
where we have introduced the dimensionless parameter $\d$.

The $R^{3/2}$-term in $F(R)$ gravity (or in the Jordan frame) arises in an approximate description of the Higgs field
with a small cubic term in its scalar potential and a large non-minimal coupling to $R$~\cite{Martins:2020oxv}.
The $R^{3/2}$ term also appears in the (chiral) modified supergravity~\cite{Ketov:2010qz,Ketov:2012se}.

The possible values of $\d$ are restricted by the condition (\ref{restr2}). Given $\tilde{\sigma}>0$, we find
\begin{equation}
\label{F32DF}
F_{,\sigma}=\fracmm{M^2_{Pl}}{6}\left(\sqrt{\tilde{\sigma}}+\fracmm{9\d}{4}\right)^2-\fracmm{3}{16}\left(27\d^2-16\right)>0
\end{equation}
when $\d>-4\sqrt{3}/9$, and
\begin{equation}
 F_{,\sigma\sigma}=\fracmm{M_{Pl}^2}{24m^2}\left(4+\fracmm{9\d}{\sqrt{\tilde{\sigma}}}\right)>0
\end{equation}
only when $\delta>0$. Hence, the condition  $\delta>0$ is necessary to get a stable $F(R)$ gravity model for all $R_J>0$.

The corresponding scalar potential (\ref{Ve}) is given by
\begin{equation}
\label{V32}
    V=\fracmm{4V_0\tilde{\sigma}(3\d\sqrt{\tilde{\sigma}}+\tilde{\sigma})}{(6+9\d\sqrt{\tilde{\sigma}}+2\tilde{\sigma})^2}
    ~~.
\end{equation}

Equation (\ref{ydef}) in this case is a quadratic equation on $\sqrt{\tilde{\sigma}}$, and its only real solution is
\begin{equation}
\label{F32sigmay}
\tilde{\sigma} = \fracmm{3(1-y)}{y} +\fracmm{9\d}{8y} \left[9\d\,y-\sqrt{3y(27\d^2y-16y+16)}\right]~.
\end{equation}

The potential (\ref{V32}) can be rewritten as
\begin{equation}
\label{Vy32}
\eqalign{    \tilde{V} & ~=\fracmm{V_0}{2304y^2}\left(s+3\d\,y\right)\left(s-9\d\,y\right)^3 \cr
& ~= \fracmm{243V_0\d^4y^2}{256} \left( 3\sqrt{1 + \fracmm{16(1-y)}{27\d^2 y}}+1\right)
\left( \sqrt{1 + \fracmm{16(1-y)}{27\d^2 y}}-1\right)^3~~, \cr}
\end{equation}
where we have introduced $s=\sqrt{3y(27\d^2y-16y+16)}$.

When $\d=4\sqrt{3}/9$, the $F_{,\sigma}$ function is a perfect square, and the potential simplifies as
\begin{equation}
\label{Vy32sd}
     \tilde{V}_{\rm special}(y)=\fracmm{V_0}{3}\left(3+\sqrt{y}\right)\left(1-\sqrt{y}\right)^3\,,
\end{equation}
or
\begin{equation*}
    V_{\rm special}(\phi)=\fracmm{V_0}{3}\left(\mathrm{e}^{\phi/(\sqrt{6}M_{Pl})}-1\right)^3\left(1+3\mathrm{e}^{\phi/(\sqrt{6}M_{Pl})}\right)\mathrm{e}^{-2\sqrt{2/3}\,\phi/M_{Pl}}\,.
\end{equation*}

\begin{figure}[htb]
\begin{subfigure}{.52\textwidth}
 \centering
 \includegraphics[width=1\linewidth]{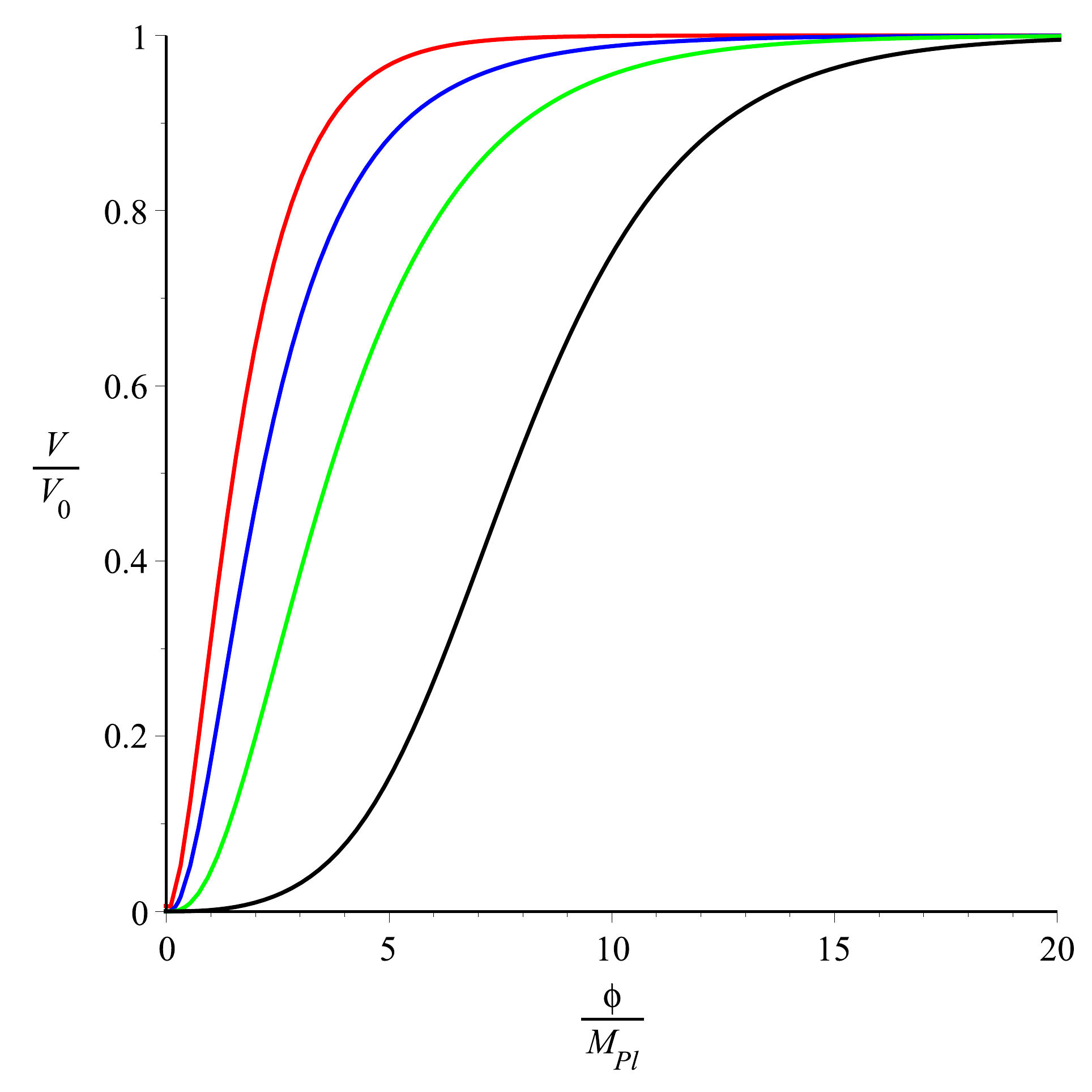}
\end{subfigure}
\caption{The potential $V(\phi)$ for $\d=0$ (red), $\d=1/5$ (blue), $\d=4\sqrt{3}/9$ (green), and $\d=5$ (black).}
\label{F32Vphi}
\end{figure}

The profile of the potential $V(\phi)$ as a function of the canonical inflaton $\phi$ is given in Fig.~\ref{F32Vphi}.

It is possible to further generalize our model by replacing the $R^{3/2}$-term by the $(R+R_0)^{3/2}$-term with a positive constant $R_0$~\cite{Ketov:2012se,Sebastiani:2013eqa}.

\subsection{The inflationary parameters}

Equations~(\ref{slr1}) and (\ref{slr2}) give the slow-roll parameters in our model (\ref{FR32}) as
\begin{equation}
\label{eps32dg}
   \epsilon=\fracmm{48y^{2} \left[s\delta+(8-9\,{\delta}^{2})y\right]^{2}}{\left(s-9\delta\,y \right)^{2}
   \left(s+3\,\delta\,y \right) ^{2}}
\end{equation}
and
\begin{equation}
\label{eta32dg}
    \eta={\fracmm{4y\left[ 24(3\delta+2)(3\delta-2)(27\delta^2-16)y^3-(27\delta(21s\delta^2-16s-80\delta)+768)y^2-s^3\delta \right] }{ \left( 3\,
\delta\,y+s \right)  \left(s-9\,\delta\,y \right) ^{2}s}}~.
\end{equation}

Therefore, according to Eq.~(\ref{nsr}), the inflationary parameters are given by
\begin{equation}
\label{nseta32dg}
\begin{split}
    n_s= ~&1+\fracmm{8y\left(3s(3\delta(9\delta^2-16)s+720\delta^2-256)y^2-s^3\delta(39\delta y+s)\right)}{ \left( -9\,\delta\,y+s \right) ^{2} \left( 3\,\delta\,y+s
 \right) ^{2}s}\\
& {}-\fracmm{8y\left[72\delta(4-9\delta^2)(27\delta^2-16)y^4+((-1215\delta^4-432\delta^2+768)s-144\delta(45\delta^2-16))y^3\right]}{ \left( -9\,\delta\,y+s \right) ^{2} \left( 3\,\delta\,y+s \right) ^{2}s}
 \end{split}
 \end{equation}
 and
\begin{equation}
\label{r32dg}
   r={\fracmm{768\,{y}^{2} \left( -9\,{\delta}^{2}y+s\delta+8\,y \right) ^{2
}}{ \left( -9\,\delta\,y+s \right) ^{2} \left( 3\,\delta\,y+s \right)^{2}}}~~.
\end{equation}

The slow-roll evolution equation (\ref{equyslr}) allows us to relate $N_e$ with $y$ at the end of inflation,
\begin{equation}
\label{Ney32dg}
\begin{split}
    N_e&=\left( {\fracmm{9}{8}}-{\delta}^{-2} \right)\ln\left[9\delta^3\left(9\delta\,y-s\right)+24\left(1 - 4y\right)\delta^2+8\left(\delta s+4y\right)\right]\\
 &{}+ \left( {\delta}^{-2}-\fracmm{3}{8}\right) \ln y  +{\fracmm{s}{4\delta\,y}}-N_0~,
\end{split}
\end{equation}
where the integration constant $N_0$ is fixed by the condition $N_e(y_{end})=0$. The analytic formula for $N_0(\d)$ is obtained by substituting $N_e=0$ and $y=y_{end}$ (see below). It results in a long equation that is not very illuminating, so we do not present it here.

It follows from Eq.~(\ref{eps32dg}) that the condition $\epsilon=1$ with arbitrary positive parameter $\d$  leads
to a quadratic equation on $y=y_{end}$ with the only solution as
\begin{equation}
\label{yend}
y_{end} = \fracmm{3(4-3\d^2+\sqrt{3}\d^2)-\sqrt{9(4-3\d^2+\sqrt{3}\d^2)^2 -72(2-3\d^2)}}{2(2-3\d^2)(3+2\sqrt{3})}~~.
\end{equation}
It is worth noticing that this solution has no singularity at $\d=\sqrt{2/3}$, while $y_{end}(\d)$ is a smooth monotonically decreasing function.

The slow-roll parameters $\epsilon$ and $\eta$ remain finite in the limit $\delta \rightarrow +\infty$ at fixed $y$,
\begin{equation}
\epsilon_\infty (y) = \fracmm{(2 y + 1)^2}{3 (1 - y)^2}~~,\quad
\eta_\infty (y) = \fracmm{2(4 y^2 + y + 1)}{3 (1 - y)^2}~.
\end{equation}

Since the value of $y$ at the end of inflation is determined by the condition $\epsilon (y_{end}) = 1$, $y_{end}$ also approaches a finite limit as $\delta \rightarrow +\infty$, which is given by a solution to the equation
\begin{equation}
\label{epsinfty}
\epsilon_\infty = \fracmm{(2 y + 1)^2}{3 (1 - y)^2} = 1.
\end{equation}
This equation has only one positive solution $y_{end}|_{\delta \rightarrow +\infty} = 3 \sqrt{3} - 5 \approx 0.196$.

The amplitude of scalar perturbations (\ref{As}) is given by
\be
A_s={\fracmm{ \left( -9\,\delta\,y+s\right)^{5}
 \left( 3\,\delta\,y+s \right)^{3}{m}^{2}}{3538944{y}^{4}{\pi }^{2} \left( -
9\,{\delta}^{2}y+s\delta+8\,y \right) ^{2}}}~~.
\ee
The observed value of $A_s$ determines the value of the parameter $m$.

\begin{figure}[t]
\begin{subfigure}{.49\textwidth}
 \centering
 \includegraphics[width=1\linewidth]{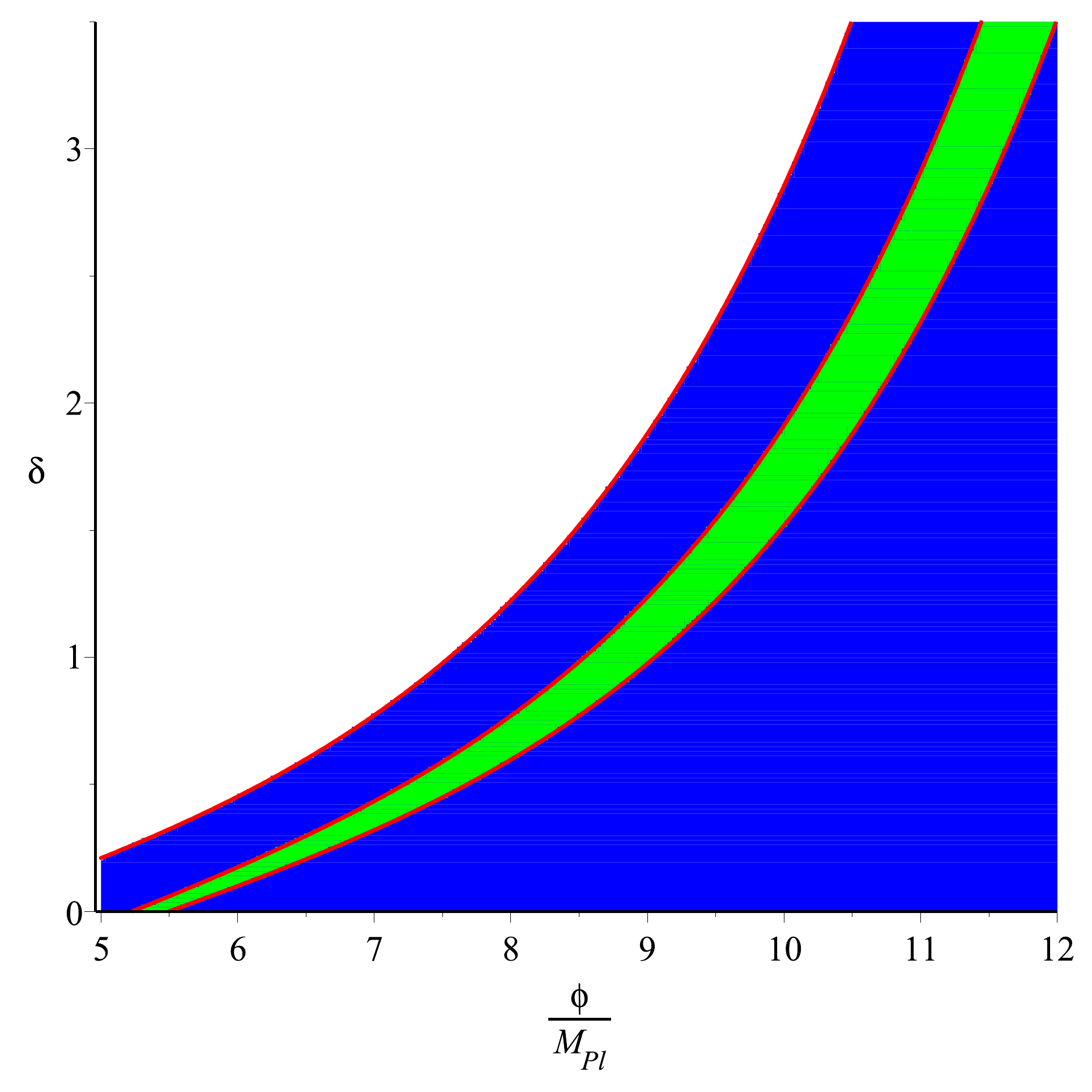}
\end{subfigure}
\begin{subfigure}{.49\textwidth}
  \centering
  \includegraphics[width=1\linewidth]{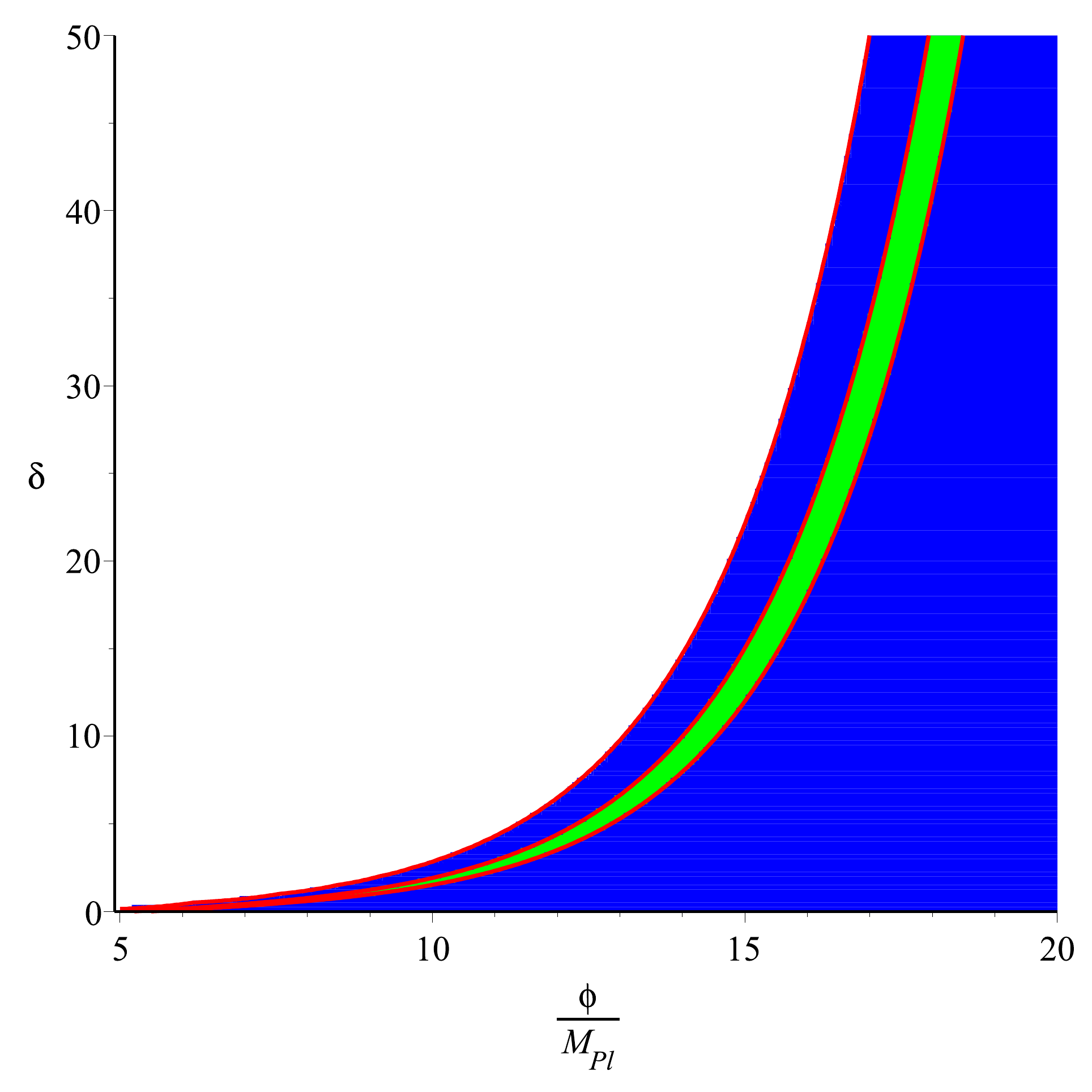}
\end{subfigure}
\captionsetup{width=.9\linewidth}
\caption{The inflaton field values against the values of the parameter $\d$. The green area corresponds to the observational restrictions on $n_s$ and $r$. The blue area is defined by the restrictions on $r$ only. When $\d>5$, the allowed domain is restricted by the lines $\d\,y^2=const$.}
\label{FigF32}
\end{figure}

We summarize our results for various values of the parameter $\d$ in Table~\ref{F32inflation} and Fig.~\ref{FigF32}.
The values of the tensor-to-scalar ratio $r$ at $\d=5$ are about four times more the corresponding values in the Starobinsky model. When $\d>5$, the $r$ is almost constant. The value of $y_{end}$ at $\d=25$ is close to its value at $\delta \rightarrow +\infty$. Unlike the models with the $R^3$ and $R^4$ terms, the potential has no extremum at $\phi<0$, so there is no fine-tuning of initial conditions. The number of e-foldings before the end of inflation monotonically increases for $\d>1$. Restricting this number to be less or equal to $65$, we get an estimation for the maximal value of the parameter as $\d=100$ at which the model still does not contradict observations.

\begin{table}[h]
\begin{center}
\caption{The values of $y$, $N_e$ and $r$ corresponding to $n_s=0.961$ and $n_s=0.969$, respectively, and the values of $y_{end}$ for some values of the parameter $\d$.\label{F32inflation}}
\begin{tabular}{|c|c|c|c|c|c|c|c|}
  \hline
  $\delta$ & $y_{end}$& ${y_{in,}}_{n_s=0.961}$ & ${y_{in,}}_{n_s=0.969}$  & ${N_{e,}}_{n_s=0.961}$ & ${N_{e,}}_{n_s=0.969}$ & $r_{n_s=0.961}$ & $r_{n_s=0.969}$ \\
   \hline
  $0$ & $0.464$ & $0.0140$ & $0.0112$ & $49.3$ & $62.3$ & $0.0043$& $0.0027$ \\
  $0.2$ & $0.395$ & $0.00682$ & $0.00505$ & $45.0$ & $56.8$ & $ 0.0096$& $ 0.0065$ \\
  $\fracmm{4\sqrt{3}}{9}$ & $0.299$ & $0.00146$ & $0.000968$ & $48.1$  & $60.9$ & $0.0152$  & $0.0099$\\
  $1$ & $0.279$ & $0.000939$ & $0.000616$ & $49.4$ & $62.4 $ & $0.0157$ & $0.0102$ \\
  $5$ & $0.205$ & $4.32\cdot10^{-5}$ & $2.75\cdot10^{-5}$ & $56.3$ & $69.7$ & $0.0168$ & $0.0108$ \\
  $10$ & $0.199$ & $1.08\cdot10^{-5}$ & $ 6.91\cdot10^{-6}$ & $58.7$ & $72.0$ & $0.0168$ & $0.0108$ \\
  $25$ & $0.197$ & $1.74\cdot10^{-6}$ & $1.11\cdot10^{-6}$ & $61.4$ & $74.8$ & $0.0169$  & $0.0108$ \\
  $50$ & $0.196$ & $4.34\cdot10^{-7}$ & $2.77\cdot10^{-7}$ & $63.5$ & $76.9$ & $0.0169$ & $0.0108$\\
  $100$ & $0.196$ & $1.09\cdot10^{-7}$ & $6.92\cdot10^{-8}$ & $65.5$ & $79.1$ & $0.0169$ & $0.0108$\\
  \hline
\end{tabular}
\end{center}
\end{table}

When $\d=4\sqrt{3}/9$, the function $F_{,\sigma}$ simplifies as
\begin{equation*}
F_{,\sigma}=\fracmm{M_{Pl}^2}{6}\left(\sqrt{\tilde{\sigma}}+\sqrt{3}\right)^2.
\end{equation*}

Accordingly, the slow-roll parameters are also simplified as
\begin{equation}
\label{eps32d}
   \epsilon=\fracmm{4y^2\left(2\sqrt{y}+y\right)^2}{3\left(\sqrt{y}-y\right)^2\left(3\sqrt{y}+y\right)^2}
\end{equation}
and
\begin{equation}\label{eta32d}
    \eta=\fracmm{4y^2\left(2\sqrt{y}+2y-1\right)}{3\left(\sqrt{y}-y\right)^2\left(3\sqrt{y}+y\right)}~.
\end{equation}

In this special case we find
\begin{equation}
\label{nseta32d}
    n_s=1-\fracmm{8y^2\left(7y+4y\sqrt{y}+y^2+3\sqrt{y}\right)}{3\left(\sqrt{y}-y\right)^2\left(3\sqrt{y}+y\right)^2}
\end{equation}
and
\begin{equation}
\label{r32d}
   r=\fracmm{64y^2\left(2\sqrt{y}+y\right)^2}{3\left(\sqrt{y}-y\right)^2\left(3\sqrt{y}+y\right)^2}~~.
\end{equation}

Equation (\ref{Ney32dg}) also simplifies to
\begin{equation}\label{Ney32d}
    N_e =\fracmm{9}{4\sqrt{y}}+\fracmm{21}{16}\ln y -\fracmm{9}{8}\ln(\sqrt{y}+2)-N_0~~,
\end{equation}
so that $\epsilon=1$ is at $y_{end} =(\sqrt{6\sqrt{3}-8}-1)^2$, and the integration constant $N_0$  is given by
\begin{equation}\label{N032d}
\begin{split}
    N_0 & =-\fracmm{1}{8}\left[21\ln(\sqrt{6\sqrt{3}-8}-1)-9\ln(1+\sqrt{6\sqrt{3}-8})+\left(6+4\sqrt{3}\right)\left(1+\sqrt{6\sqrt{3}-8}\right)\right] \\
&    \approx -1.48~~.
\end{split}
\end{equation}

The inflationary parameters in the special case are also given in Table~\ref{F32inflation}. It is worth noticing that the values of the tensor-to-scalar ratio $r$ are significantly higher than those in the Starobinsky model.

The amplitude of scalar perturbations at the special value of $\d$ is given by
\be
A_s\approx\fracmm{2  \tilde{V}}{3\pi^2M_{Pl}^4 r}\approx\fracmm{ m^2 (1-\sqrt{y})^5(3+\sqrt{y})^3}{128\pi^2 M_{Pl}^2 (2+\sqrt{y})^2}~~.
\ee
Its observed value  $A_s\approx 2.1\cdot 10^{-9}$ implies $m\approx 2\cdot 10^{-5}$.

\section{Deforming the scalar potential in the Starobinsky model with analytic $F$-functions}

The Starobinsky inflation based on the inflaton potential (\ref{starp}) is a {\it large} $\phi$-field inflation, with the relevant
values of $\phi$ around the Planck scale. Hence, the field $y$ defined by (\ref{ydef})
is {\it small\/} during slow-roll inflation. The inflaton potential (\ref{starp}) as a function of $y$ is
\be  \lb{starcmb}
V(\phi)=\tilde{V}(y)= V_0 \left[ 1- 2y +{\cal O}\left(y^2\right)\right]~,
\ee
 where only the first two terms are essential for the CMB observables~\cite{Nakada:2017uka}.
The inflaton potential (\ref{starp}) can therefore be modified as
\be \lb{ano}
\tilde{V}(y)= V_0 \left[ 1- 2y +y^2\o(y)\right]\,,
\ee
with arbitrary analytic function $\o(y)$ {\it without} changing the CMB observables predicted by the
Starobinsky model, at least for those values of $\o$ that are not very large. The Starobinsky model appears at $\o=1$ with inflation taking place for positive values of $R_J$.

The physical conditions (\ref{restr2}) are satisfied with the potential (\ref{ano}). In particular, we find the second derivative $F_{,\s\s}(y)$ in the form
\be
\lb{Fsndder}
F_{,\s\s}(y) = \fracmm{M^2_{Pl}}{3 m^2 \left(2 + 2y^3 \fracmm{d \o}{d y} + y^4 \fracmm{d^2 \o}{d y^2}\right)}~~.
\ee

Equation (\ref{Ry}) reads
 \be \lb{ry1}
\tilde{\sigma}\equiv\fracmm{R_J}{m^2}= 3\left(\fracmm{1}{y}  -1 - \fracmm{1}{2} y^2 \fracmm{d\o}{dy}\right)~~,
 \ee
 and Eq.~(\ref{Fy}) is given by
 \be \lb{fy1} F = V_0\left(\fracmm{1}{y^2} -\o -y\fracmm{d\o}{dy}\right)~.
 \ee
 As a check, in the Starobinsky case, $\o=1$ and $V=V_0(1-y)^2$, and Eq.~(\ref{Ry}) gives
 \be \lb{starinv}
 y = \left( 1+ \fracmm{R_J}{3m^2}\right)^{-1}~.
 \ee
 Substituting it into Eq.~(\ref{Fy}), we get
 \be
 F_{Star.}(R_J) = \fracmm{M^2_{Pl}}{2}\left( R_J + \fracmm{R_J^2}{6m^2}\right)~,
 \ee
 as it should be. Moreover, when $\o$ is an arbitrary constant, we find
  \be \lb{starc}
  F(R_J) = F_{Star.}(R_J) -\L~,
 \ee
 where $\L=V_0(1-\o)$ is a cosmological constant.

\subsection{New case I}

As a new example, let us now consider the non-trivial case with
\be \lb{olin}
\o(y) = \o_0 + \o_1 y ~,
\ee
where $\o_0 \leqslant 1$ and $\o_1>0$ are constants. The constant $\o_1$ should be positive for the potential $V$  bounded from below. The inequality $\o_0 \leqslant 1$ is needed for positivity of a cosmological constant, see
Eq.~(\ref{starc}).

\begin{figure}[htb]
\begin{subfigure}{.49\textwidth}
 \centering
 \includegraphics[width=1\linewidth]{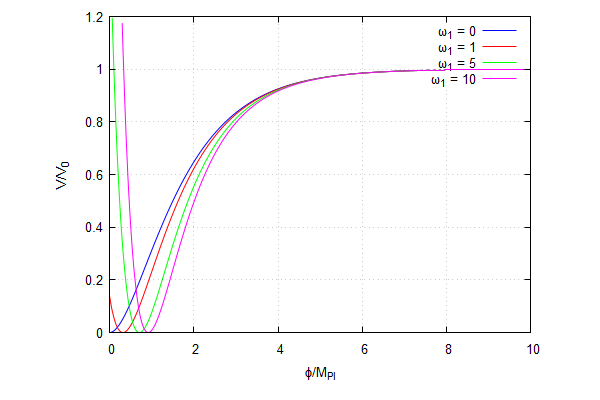}
\end{subfigure}
\begin{subfigure}{.49\textwidth}
  \centering
  \includegraphics[width=1\linewidth]{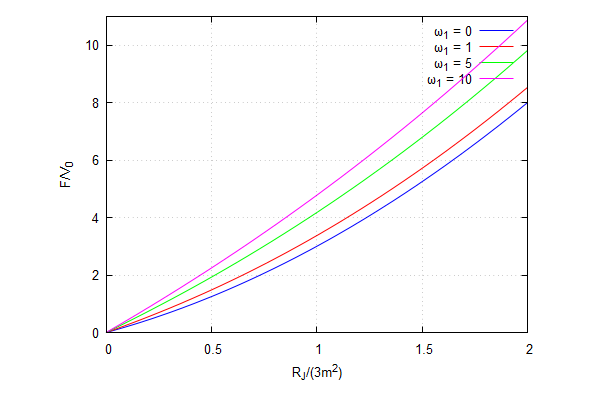}
\end{subfigure}
\captionsetup{width=.9\linewidth}
\caption{The scalar potential of the canonical inflaton field $\phi$ (left) and the related $F(R)$-function (right)
in the case of the deformation of the Starobinsky model for some values of the parameter $\o_1$: $0$, $1$, $5$, and $10$.}
\label{Fig_cubic}
\end{figure}

Equation (\ref{ry1}) leads to the depressed cubic equation
\be \lb{depcub}
y^3 + \fracmm{2}{\o_1}\left( 1 + \fracmm{R_J}{3m^2}\right)y - \fracmm{2}{\o_1} \equiv y^3+py+q=0
\ee
with the negative discriminant
\be \lb{discr}
\D ={} -\left(4p^3+27q^2\right)= -\fracmm{32}{\o^3_1}\left( 1 + \fracmm{R_J}{3m^2}\right)^3 - \fracmm{108}{\o^2_1} <0~,
\ee
so that it has only one real root given by the Cardano formula
\be \lb{cardano1}
\eqalign{
y  & = \left( -\fracmm{q}{2} +\sqrt{\fracmm{q^2}{4} +\fracmm{p^3}{27}}\right)^{1/3} +
\left( -\fracmm{q}{2} -\sqrt{\fracmm{q^2}{4} +\fracmm{p^3}{27}}\right)^{1/3} \cr
 &  = \fracmm{1}{\o_1^{1/3}} \left( 1 +\sqrt{ 1+\fracmm{8}{27\o_1} \left( 1 + \fracmm{R_J}{3m^2}\right)^3} \right)^{1/3}
+ \fracmm{1}{\o_1^{1/3}} \left( 1 -\sqrt{ 1+\fracmm{8}{27\o_1} \left( 1 + \fracmm{R_J}{3m^2}\right)^3} \right)^{1/3}~.}
\ee
Equation (\ref{fy1}) yields the explicit $F$-function in this case as follows:
\be \lb{case1}
\eqalign{
\fracmm{F}{V_0} & = \fracmm{1}{y^2} -\o_0 - 2\o_1y \cr
& = \o_1^{2/3} \left[
\left( 1 +\sqrt{ 1+\fracmm{8}{27\o_1} \left( 1 + \fracmm{R_J}{3m^2}\right)^3} \right)^{1/3}
+ \left( 1 -\sqrt{ 1+\fracmm{8}{27\o_1} \left( 1 + \fracmm{R_J}{3m^2}\right)^3} \right)^{1/3}
\right]^{-2} \cr
& - 2\o_1^{2/3} \left[
\left( 1 +\sqrt{ 1+\fracmm{8}{27\o_1} \left( 1 + \fracmm{R_J}{3m^2}\right)^3} \right)^{1/3}
+ \left( 1 -\sqrt{ 1+\fracmm{8}{27\o_1} \left( 1 + \fracmm{R_J}{3m^2}\right)^3} \right)^{1/3}
\right] \cr
& - \o_0~,}
\ee
where
\be
\eqalign{
\o_0 &= \o_1^{2/3} \left[
\left( 1 +\sqrt{ 1+\fracmm{8}{27\o_1}}\right)^{1/3}
+ \left( 1 -\sqrt{ 1+\fracmm{8}{27\o_1}}\right)^{1/3}
\right]^{-2} \cr
& - 2\o_1^{2/3} \left[
\left( 1 +\sqrt{ 1+\fracmm{8}{27\o_1}}\right)^{1/3}
+ \left( 1 -\sqrt{ 1+\fracmm{8}{27\o_1}} \right)^{1/3}
\right]~.}
\ee
 The limit $\o_1\to 0$ is smooth and gives back the case (\ref{starc}). The shape of the scalar potential
 of the canonical inflaton field $\phi$ is shown on the left-hand-side of Fig.~\ref{Fig_cubic}: it remains largely unchanged  for the values of the deformation parameter up to $\o_1\approx 10$,  though the "waterfall" down to the minimum  becomes steeper and the non-vanishing vacuum expectation value $\VEV{\phi}\neq 0$ appears for $\o_1>0$.  Since a vacuum expectation value should be less than $M_{Pl}$~, we find that the deformation parameter
 $\o_1 < 2 \left(1 -e^{-\sqrt{2/3}}\right)e^{3\sqrt{2/3}}\approx 12.9 $.

 The corresponding $F(R)$ gravity functions are shown on the right-hand-side of Fig.~\ref{Fig_cubic}. Compared to the Starobinsky curve $(\o_1=0)$, they have larger values in the given range of $R_J/m^2$. For larger values of
 $R_J/m^2$, all those $F$-curves converge to the Starobinsky curve.

\begin{figure}[htb]
\begin{subfigure}{.49\textwidth}
 \centering
 \includegraphics[width=1\linewidth]{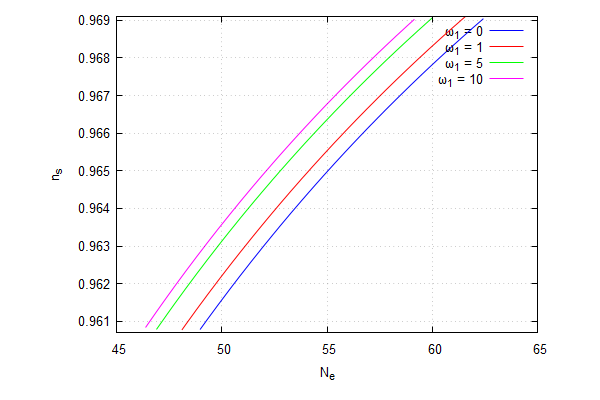}
\end{subfigure}
\begin{subfigure}{.49\textwidth}
  \centering
  \includegraphics[width=1\linewidth]{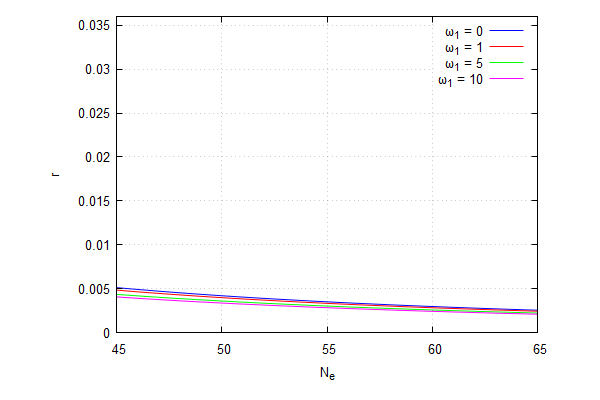}
\end{subfigure}
\captionsetup{width=.9\linewidth}
\caption{The index $n_s$ (left) of scalar perturbations and the tensor-to-scalar ratio $r$ (right) as the functions
of e-folds $N_e$ in the case I of the deformation of the Starobinsky model for some values of the parameter $\o_1$: $0$, $1$, $5$, and $10$.}
\label{Fig_cubicTilts}
\end{figure}

The results of our numerical calculation for the index $n_s$ of scalar perturbations and the tensor-to-scalar ratio $r$,
 as the functions of e-folds $N_e$, are given in Fig.~\ref{Fig_cubicTilts} for the values of the parameter $\o_1$ equal to $0$, $1$, $5$, and $10$. As is clear from Fig.~\ref{Fig_cubicTilts}, there are (small) changes in the duration of inflation and its initial and final moments.

From Eq.~(\ref{Fsndder}) we get a simple formula for the second derivative $F_{,\s\s}$ as
\be
F_{,\s\s}(y) = \fracmm{M^2_{Pl}}{6 m^2 \left(1 + \o_1 y^3\right)}~~,
\ee
so that the $F(R)$ gravity model (\ref{case1}) satisfies the conditions (\ref{restr2}) with $\o_1\geqslant 0$.

\subsection{New case II}

Our approach suggests another one-parametric deformation of the Starobinsky potential (\ref{starp}) as follows:
\be  \lb{pot2}
V(\phi) = V_0 \left[ 1 - e^{-\sqrt{2/3}\phi/M_{Pl}} -\z e^{-2\sqrt{2/3}\phi/M_{Pl}}\right]^2=V_0\left(1-y-\z y^2\right)^2\,,
\ee
where we assume the parameter $\z\geqslant 0$. This potential can be realized in supergravity~\cite{Ketov:2019toi},
while the potential (\ref{starp}) is recovered at $\z=0$.
The Minkowski minimum of the potential (\ref{pot2}) is realized by demanding $V(y)=V'(y)=0$ that yields
a quadratic equation whose solution is given by
\be \lb{min2}
y_{\rm min.} = \fracmm{\sqrt{1+4\z} -1}{2\z}~.
\ee
The potential (\ref{pot2}) can also be rewritten to the form (\ref{ano})
with the $\o(y)$-function
\be \lb{omega2}
\o = 1-2\z + 2\z y +\z^2y^2~.
\ee
Accordingly, Eq.~(\ref{ry1}) takes the form
\be \lb{quartic2}
\z^2y^4+\z y^3 +\left( 1 + \fracmm{R_J}{3m^2}\right)y - 1 \equiv ay^4+by^3+cy^2+dy +e=0
\ee
with
\be \lb{qdata}
a=\z^2,\quad b=\z,\quad c=0,\quad d=1+\fracmm{R_J}{3m^2},\quad e={}-1~.
\ee
The discriminant of the quartic equation (\ref{quartic2}),
\be
\lb{qdiscr}
\begin{split}
\D &={} -256 \z^6 - 192\z^5\left( 1+ \fracmm{R_J}{3m^2}\right)
-3\z^4 \left[ 9\left( 1+ \fracmm{R_J}{3m^2}\right)^4- 2\left( 1+ \fracmm{R_J}{3m^2}\right)^2+9\right]\\
&{}-4\z^3\left( 1+ \fracmm{R_J}{3m^2}\right)^3,
\end{split}
\ee
is negative, so that there are two distinct real roots and two conjugated complex roots.
The equivalent depressed quartic equation
\be \lb{deprq}
\tilde{y}^4 + p\tilde{y}^2 + q\tilde{y} +r =0,\quad {\rm where} \quad \tilde{y}=y+\fracmm{b}{4a}=y+\fracmm{\z^{-1}}{4}~~,
\ee
has the following parameters:
\be \lb{deprqp1}
p = {}-\fracmm{3b^2}{8a^2} ={}- \fracmm{3}{8\z^2} <0~,\quad q=\fracmm{b^3+8a^2d}{8a^3}=
\fracmm{1+8\z\left( 1+ \fracmm{R_J}{3m^2}\right)}{8\z^3}>0~,
\ee
and
\be \lb{deprqp2}
r = {}-\fracmm{3 + 256\z^2 +64\z\left( 1+ \fracmm{R_J}{3m^2}\right)}{256\z^4}<0~~.
\ee
The Ferrari formula gives the roots of the depressed quartic equation in terms of the quantities
\be \lb{quarticD}
\eqalign{
\D_0 & =-3bd + 12ae = -3\z\left( 1+ \fracmm{R_J}{3m^2}\right)-12\z^2 <0~~,\cr
\D_1 & = 27b^2e + 27ad^2 = 27\z^2 \left[ \left( 1+ \fracmm{R_J}{3m^2}\right)^2-1\right] >0~,\cr
Q & = \left( \fracmm{\D_1+\sqrt{\D_1^2-4\D^3_0}}{2}\right)^{1/3}~,\quad
S = \fracmm{1}{2} \sqrt{ -\fracmm{2}{3}p + \fracmm{1}{3a}\left( Q + \fracmm{\D_0}{Q}\right)}~,}
\ee
where in our case $Q>0$, $S>0$ and $27\D=4\D_0^3-\D^2_1 <0$. The real roots are given by
\be \lb{2roots}
y_{1,2}  = -\fracmm{b}{4a}-S \pm \fracmm{1}{2} \sqrt{-4 S^2-2p+\fracmm{q}{S}}
 = -\fracmm{1}{4\z} -S\pm \fracmm{1}{2} \sqrt{-4 S^2 +\fracmm{3}{4\z^2}
 +\fracmm{1+8\z\left( 1+ \fracmm{R_J}{3m^2}\right)}{8S\z^3}}~.
\ee
 In the limit $\z\to 0_+$ we find
\begin{equation*}
 p\to{}-\fracmm{3}{8\z^2}~,\quad q\to \fracmm{1}{8\z^3}~,\quad \D_1\to 0~,\quad \D_0\to -3\z\left( 1+ \fracmm{R_J}{3m^2}\right)~,\quad Q\to \sqrt{\abs{\D_0}}~,\quad S\to \fracmm{1}{4\z}~.
\end{equation*}

\begin{figure}[htb]
\begin{subfigure}{.49\textwidth}
 \centering
 \includegraphics[width=1\linewidth]{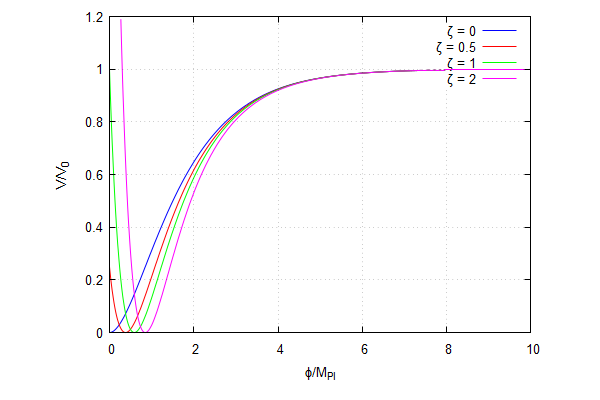}
\end{subfigure}
\begin{subfigure}{.49\textwidth}
  \centering
  \includegraphics[width=1\linewidth]{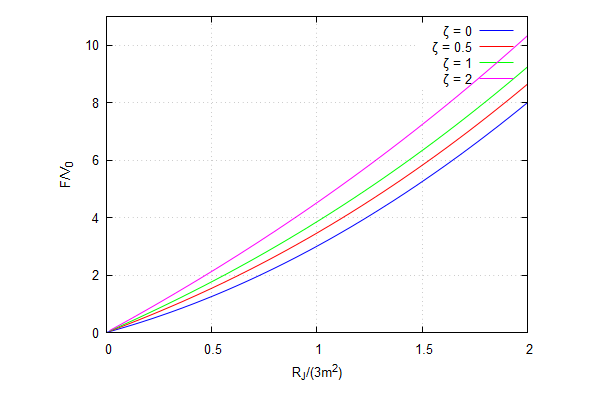}
\end{subfigure}
\captionsetup{width=.9\linewidth}
\caption{The scalar potential of the canonical inflaton field $\phi$, and the related $F(R)$ function
in case II.}
\label{Fig_quartic}
\end{figure}

Actually, only the root with the {\it upper sign} choice in Eq.~(\ref{2roots}) leads to the physical solution connected to the Starobinsky model because the lower sign choice implies the negative value of $y$ and, therefore, should be
discarded. The scalar potential and the $F$ function are given in Fig.~\ref{Fig_quartic} for some
values of the deformation parameter $\z$. Since a vacuum expectation value  should be less than $M_{\rm Pl}$~, we find a restriction on the deformation parameter,
\be
\lb{zetarest}
\z < \left(1 -e^{-\sqrt{2/3}}\right)e^{2\sqrt{2/3}}\approx 2.9 ~.
\ee

\begin{figure}[htb]
\begin{subfigure}{.49\textwidth}
 \centering
 \includegraphics[width=1\linewidth]{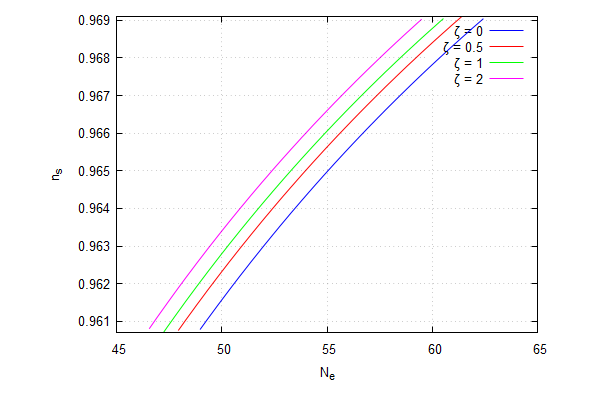}
\end{subfigure}
\begin{subfigure}{.49\textwidth}
  \centering
  \includegraphics[width=1\linewidth]{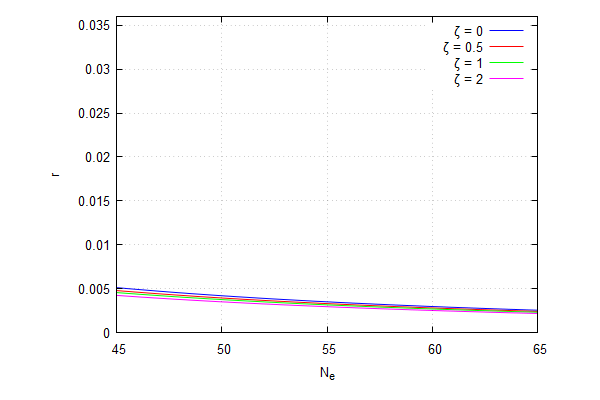}
\end{subfigure}
\captionsetup{width=.9\linewidth}
\caption{The index $n_s$ (left) of scalar perturbations and the tensor-to-scalar ratio $r$ (right) as the functions
of e-folds $N_e$ in case II for some values of the parameter $\z$: $0$, $1/2$, $1$, and $2$.}
\label{Fig_quarticTilts}
\end{figure}

The results of our numerical calculation for the index $n_s$ of scalar perturbations and the tensor-to-scalar ratio $r$,
 as the functions of e-folds $N_e$, in case II are given in Fig.~\ref{Fig_quarticTilts} for the values of the parameter $\z$ equal to $0$, $1/2$, $1$, and $2$. There are (small) changes in the duration of inflation and its initial and final moments, similarly to case I.

From Eq.~(\ref{Fss}), we find
\begin{equation*}
F_{,\s\s}(y) = \fracmm{M^2_{Pl}}{6 m^2 \left(1 + 2 \z y^3 + 3 \z^2 y^4\right)}>0~~.
\end{equation*}

To the end of this Section, we comment on case II with a negative parameter $\z=-\abs{\z}$. The critical points of the
potential (\ref{pot2}) correspond to three solutions,
\be \lb{crpot2}
y_{1,2}=\fracmm{1\mp \sqrt{1-4\abs{\z}}}{2\abs{\z}}~,\quad y_3=\fracmm{1}{2\abs{\z}}~.
\ee
To get slow-roll inflation, we need $\abs{\z}<1/4$. It leads to the scalar potential with two minima and
one maximum, while only one minimum  has $y<1$, thus leading to hilltop inflation with restricted initial conditions.

\section{Conclusion}

In this paper, we studied several extensions of the Starobinsky inflation model of the $(R+R^2)$ gravity in the context of $F(R)$ gravity and scalar-tensor gravity, including the known models (Sec.~3) and the new ones (Secs.~4 and 5) by
using our methods. These models are consistent with cosmological observations and have no ghosts, while {\it both} the $F$-function and the inflaton (scalaron) potential $V(\phi)$ are available in the explicit analytic form including their dependence upon the parameters. The last feature appears to be very  restrictive. We focused on the models with only one free parameter for simplicity. Of course, the list of such models can be extended but not much. Besides the physical and observational constraints, on the one side, it is must be possible to invert the dependence of a polynomial in $R$ upon the inflaton field $\phi$ as an analytic function $R(\phi)$, see Sec.~3. On the other side, it must be possible to analytically invert a function $R(y)$, see Sec.~5. It is only possible when there is an analytic formula for the roots of the polynomials, which restricts their order to 4 or less.

In Secs.~3 and 4 we deformed the Starobinsky $(R+R^2)$ gravity model of inflation and found the inflaton potential in the analytic form in the three specific cases, by adding an $R^3$ term, an $R^4$-term and an $R^{3/2}$ term, respectively. In Sec.~5 we started from the opposite side, by deforming the scalar potential of the Starobinsky model, and derived the corresponding $F$-function in the analytic form in the two different models with a single parameter, and found the lower and upper bounds on the values of the parameters. The new models in Sec.~5 are very close to the original Starobinsky model of inflation, as regards their cosmological parameters. However, unlike the Starobinsky model, the inflaton (scalaron) acquires a non-vanishing vacuum expectation value in those models.

The asymptotic de Sitter solution in the pure $(R+R^2)$ model at very large positive values of $R$  turns out to be unstable against any correction of the higher order in $R$ in the $F$-function that drastically changes the behavior of the scalar potential before inflation. Even though those values of $R$ are apparently beyond the scope of applicability of the modified $F(R)$-gravity as the effective  theory of gravity, it introduces a dependence upon initial values of the inflaton field that must roll down to the left from the maximum of the potential (see Figs.~\ref{Fig_cubic} and \ref{Fig_quartic}) for viable hilltop inflation. In other words, the higher-order (in $R$) quantum corrections imply tuning the initial conditions for inflation.~\footnote{The similar results were obtained in the $(R+R^2+R^3)$
model~\cite{Rodrigues-da-Silva:2021jab}.} This feature is apparently universal because there is no fundamental reason for the absence of quantum corrections proportional to the higher powers of $R$. This observation is, however, limited to the perturbative treatment with respect to the $R$-dependence of $F$-function.

Consistency with CMB observations in our models demands the higher order terms in $R$ to be negligible against
the $R^2$ term during inflation. We achieved it via demanding the dimensionless coefficients in front of the higher-order terms to be small enough because the variable $\tilde{\s}=R/m^2$ is not small during inflation. We found the
upper limits on the values of the coefficients at the $R^3$ and $R^4$ terms. However, we are not aware of any fundamental reason demanding those coefficients to be small.~\footnote{It might be possible to get a flat potential for very large inflaton field values via a resummation of all higher-curvature corrections, e.g., in the framework of asymptotically-safe quantum gravity \cite{Liu:2018hno,Chojnacki:2021fag}. }

Unlike the $R^3$ and $R^4$ terms, the modification of the Starobinsky model by the $R^{3/2}$ term does not lead
to significant constraints on its coefficient in slow-roll inflation, at least for $0<\d<100$. However, the
$R^{3/2}$ term has a significant impact on the value of the tensor-to-scalar ratio $r$ (see Table 1 in Sec.~4)~\footnote{It also has a significant impact on reheating after inflation~\cite{Ketov:2012se}.} and, therefore, implies higher production rate of primordial gravitational waves caused by inflation. There is another source of primordial gravitational waves caused by possible formation of primordial black holes; in the framework of $F(R)$ gravity it was studied in Ref.~\cite{Papanikolaou:2021uhe}.

More restrictions of the $F$-function and the inflaton scalar potential arise when demanding their minimal embedding
into supergravity, along the lines of Refs.~\cite{Ketov:2010qz,Farakos:2013cqa,Ferrara:2013rsa}. For instance, the
$R^3$ term is excluded in supergravity, whereas only case II in Subsec.~5.2 is extendable in the minimal supergravity framework that requires the scalar potential to be a real function squared. The $R^{3/2}$ term also arises
in certain versions of the chiral $F({\cal R})$ supergravity \cite{Ketov:2010qz}.

\acknowledgments

The authors are grateful to Gia Dvali, Olaf Lechtenfeld, Kei-ichi Maeda and Alexei Starobinsky  for discussions and correspondence.

V.R.I. was supported by the Theoretical Physics and Mathematics Advancement Foundation ``BASIS''.
S.V.K. was supported by Tokyo Metropolitan University, the World Premier International Research Center Initiative (MEXT) and Yamada Foundation in Japan.
S.V.K. is grateful to Leibniz Universit\"at Hannover in Germany for pleasant hospitality during this investigation.
E.O.P. and S.Yu.V. were partially supported by the Russian Foundation for Basic Research grant No.~20-02-00411.

\bibliography{Bibliography}{}
\bibliographystyle{JHEP}
\end{document}